\documentclass[12pt]{article}   

\usepackage{graphicx}
\usepackage{amsmath}    
\usepackage{amsfonts}
\usepackage{epsfig}

\begin{document}

\title{\LARGE{
{MEtop - a generator for single top production via FCNC interactions}
}}

\author{
\large
Rita Coimbra,$^{(1)}$\thanks{E-mail: rita.coimbra@coimbra.lip.pt} \
Ant\'onio Onofre,$^{(2)}$\thanks{E-mail: Antonio.Onofre@cern.ch}\
Rui Santos,$^{(3,4)}$\thanks{E-mail: rsantos@cii.fc.ul.pt} \
and Miguel Won$\, ^{(1)}$\thanks{E-mail: miguel.won@coimbra.lip.pt}
\\*[5mm]
\small $^{(1)}$ LIP / Departamento de F\'{\i}sica,  \\
\small Universidade de Coimbra, 3004-516 Coimbra, Portugal
\\*[2mm]
\small $^{(2)}$  Departamento de F\'{\i}sica, \\
\small Universidade do Minho, 4710-057 Braga, Portugal
\\*[2mm]
\small $^{(3)}$ Instituto Superior de Engenharia de Lisboa \\
\small 1959-007 Lisboa, Portugal
\\*[2mm]
\small $^{(4)}$ Centro de F\'\i sica Te\'orica e Computacional,
Universidade de Lisboa  \\
\small 1649-003 Lisboa, Portugal
}

\date{\today}

\maketitle

\begin{abstract}
We present a generator for single top quark production via
flavour-changing neutral currents. The MEtop event generator
allows for Next-to-Leading-Order direct top production $pp \to t$
and Leading-Order production of several other single top
processes.  A few packages with definite sets of dimension six operators  
are available. We discuss how to improve the bounds on 
the effective operators and how well new physics can be probed
with each set of independent dimension six operators. 
\end{abstract}

\newpage

\section{Introduction}

The Large Hadron Collider (LHC) at CERN has resumed operation with its 
center-of-mass energy increased to 8 TeV. The LHC top factory will allow
us to scrutinise the heaviest of all known quarks with unprecedented precision. 
Flavour physics is on the top of the agenda as one of most interesting research topic that can be 
address at this collider, through the study of flavour changing neutral currents 
(FCNC) in top quark production and decay. In fact, a wide variety of models
shows a strong dependence in the measurable FCNC quantities: for instance, 
top quark FCNC branching ratios can vary from extremely small in the
Standard Model (SM) to measurable values at the LHC
in a variety of the SM extensions~\cite{AguilarSaavedra:2004wm}. Therefore, the
large number of top quarks being produced provide a means to search for deviation 
from the SM, however small they are. It is clear that the simplest way
to search for new FCNC physics is to look for the rare top decays as 
for example in $t \to q \gamma$ where $q=u,c$ is the up-quark
or a c-quark. Limits on the  
$BR(t \to q \gamma)$ were set indirectly at LEP~\cite{LEP2Zgamma} 
and HERA\cite{Abramowicz:2011tv, Aaron:2009vv} and
directly at the Tevatron\cite{gammaCDF}  (see~\cite{Guedes:2010wr}
for references and details). Presently, the best bound on the photon
FCNC current is the one from HERA while the best experimental
bounds on $BR(t \to q Z)$ were obtained at the Tevatron~\cite{Abazov:2011qf} 
and at the LHC~\cite{Aad:2012ij, CMStqZ} . Finally, the best bound on the strong FCNC
current $tqg$ was recently obtained in direct top production at the LHC~\cite{Aad:2012gd}.

Our goal is to provide a tool to measure
FCNC related to the top quark at the  production level.
More evolved top FCNC searches can only be performed if a dedicated generator
for top FCNC studies is available. So far only the PROTOS generator~\cite{PROTOS}
and the TopReX generator~\cite{Slabospitsky:2002ag}
can be used to generate events for direct top production $pp (g u + g c) \to t$
studies as well as the top FCNC decays. Therefore, we considered that 
it was both necessary and timely
to make available a generator for top FCNC physics that included  a larger
set of FCNC operators together with a larger set of physical processes
at the production level. This
is the purpose of the MEtop event generator.

MEtop main process is direct top production, $pp (g u + g c) \to t$, 
but calculated at Next-to-Leading-Order (NLO). NLO direct top
was implemented by adopting an effective NLO approximation as described
in~\cite{Boos:2006af}. Besides direct top, MEtop can be used to generate
events at LO for all FCNC processes with a top and a gluon or  any quark
other then the top in the final state. We plan to include other processes 
like $pp \to t V$, with $V=\gamma, Z, W$ in the near future. 
From the theoretical point of view we will adopt the effective operator formalism
as described in~\cite{buch}. We use a set of dimension six effective operators
always involving at least one top quark. The set of operators
is classified in three different groups: strong, electroweak and
four-fermion (4F) operators. MEtop comes with several choices of packages
where different sets of effective operators are available.

The plan of the paper is as follows. The next section 
describes the complete set of FCNC operators needed
for top physics. In section 3 we provide a detailed description
of the physical processes available in MEtop. Section 4 is devoted
to the implementation in MEtop of the effective NLO approximation
for direct top production. In section 5 we compare the contributions
of the different classes of operators to single top production. Our conclusions
are drawn in section 6. There are also three appendixes dealing
with more technical issues.

\section{The FCNC operators for top physics}

The effective operator formalism assumes that some general
theory which has the SM as its low energy limit can be written
as a series in  $\Lambda$ with operators obeying  
the SM symmetries,
\begin{equation}
{\cal L} \;\;=\;\; {\cal L}^{SM} \;+\; \frac{1}{\Lambda}\,{\cal
L}^{(5)} \;+\; \frac{1}{\Lambda^2}\,{\cal L}^{(6)} \;+\;
O\,\left(\frac{1}{\Lambda^3}\right) \;\;\; , \label{eq:l}
\end{equation}
where ${\cal L}^{SM}$ is the SM lagrangian and ${\cal L}^{(5)}$ and 
${\cal L}^{(6)}$ contain all the dimension five and six operators respectively.
This formalism provides a model-independent parametrisation of physics
beyond the SM. This lagrangian contains
only SM fields and therefore any new particles and any new interactions
are hidden in the effective operators. The term ${\cal L}^{(5)}$ vanishes
if baryon and lepton number conservation is imposed.

The complete set of dimension six operators is quite vast. 
In order to simplify the discussion we classify the operators in three categories:
strong FCNC  operators~\cite{Ferreira:2005dr, Ferreira:2006xe}, 
the ones generating a vertex of the form $\bar t u g$, where $g$
is a gluon and $u$ is an up-quark; electroweak FCNC 
operators~\cite{Ferreira:2008cj, Coimbra:2008qp}, which
are the ones giving rise to a vertex with one top quark, an up-quark 
and one electroweak gauge boson, and finally four fermion  (4F) operators
which are Fermi interactions with one top quark and three other
quarks. 

When writing all allowed dimension six operators obeying
the required symmetries of the SM lagrangian, one readily understands
that not all operators are independent~\cite{buch}. They are related by 
the equations of motion and also by Fierz transformations.  Therefore,
the total number of operators can be reduced to a minimum set of independent operators.
Moreover, this set can be further reduced when only specific
processes are studied, like in our case, where all operators have at least
one top quark in the interaction. A minimal set of operators for top quark
physics was discussed in~\cite{Ferreira:2005dr, Ferreira:2006xe, AguilarSaavedra:2008zc, Grzadkowski:2010es}
and here we will just present this minimal set according to our classification.

We will start by considering the non 4F operators. As previously stated, we divide
these operators in two classes: strong FCNC operators, when the gluonic tensor
is involved, and electroweak FCNC operators when electroweak gauge bosons
are present in the interaction. We assume that 
$\mathcal O^{ij}$ and $\mathcal O^{ji}$ are independent operators and the 
hermitian conjugate of all the operators are included in the final lagrangian.

Following the notation of~\cite{buch}, the independent operators contributing to the strong 
FCNC vertices can be written as
\begin{equation}
{\cal O}^{ij}_{uG \phi} \, = \,  \bar{q}^{i}_L \, \lambda^{a} \, \sigma^{\mu\nu} \, u^{j}_R \, \tilde{\phi} \, G^{a \mu \nu}  \,  ,
\label{eq:op1}
\end{equation}
where $G^a_{\mu\nu}$ is the gluonic field tensor, $u^i_R$ stands for a 
right-handed quark singlet and $q^i_L$ represents the left-handed quark 
doublet. FCNC occurs because one of the indices is always equal to 3 while the 
other is either 1 or 2, that is, there is always one (and one only) top-quark 
present in the operator; the remaining fermion field in the interaction is 
either a u or a c-quark.  These operators will give rise to the FCNC vertices of the 
form $g\,t\,\bar{u_i}$ (with $u_i  = u, \, c$) and the corresponding 
hermitian conjugate interaction with an independent coefficient. Operator
in equation (\ref{eq:op1}) also appears in the literature as a dimension 5 operator. In that
case, the corresponding FCNC lagrangian is written as
\begin{equation}
{\cal L_S}  = i \kappa_u \, \frac{g_s}{\Lambda} \bar{u} \lambda^a \sigma^{\mu\nu} (f_u+h_u \, \gamma_5) t  G_{\mu\nu}^a \quad + (u \leftrightarrow c) + h.c. 
\end{equation}
where $\kappa_u$ is real, $g_s$ is the strong coupling and $f_u$ and $h_u$ are complex numbers with $|f_u|^2 + |h_u|^2 = 1$ (see appendix A for a detailed
discussion relating the forms of the strong FCNC operators).

In the electroweak sector we now have to look for FCNC vertices of the type 
$V  \, t  \, \bar{u_i}$ (with $u_i  = u, \, c$ and $V = Z, \gamma$). The minimal
set of operators that give rise to the above interactions can be written as  
\begin{equation}
{\cal O}^{ij}_{uB \phi}  \, = \, \bar{q}^i_L \, \sigma^{\mu\nu} \, u^j_R \, \tilde{\phi} \,B_{\mu\nu} \, , \quad
{\cal O}^{ij}_{uW \phi} \, = \, \bar{q}^i_L \, \tau_{I} \, \sigma^{\mu\nu} \, u^j_R \, \tilde{\phi} \, W^{I}_{\mu\nu} \,  ,
\label{eq:op3}
\end{equation}
\\[-0.3cm]
\begin{equation}
{\cal O}^{ij}_{\phi u}  \, = \, i  \, (\phi^{\dagger}  D_{\mu} \phi) \, (\bar{u}^i_R \,  \gamma^{\mu}  \,  u^j_R) \, , 
\\
\label{eq:op4}
\end{equation}
\\[-0.3cm]
\begin{equation}
{\cal O}^{(1),ij}_{\phi q} \, = \, i \,  (\phi^{\dagger} D_{\mu} \phi) \, (\bar{q}^i_L   \, \gamma^{\mu}  \,  q^j_L)  \, , \quad
{\cal O}^{(3),ij}_{\phi q} \, = \, i \, (\phi^{\dagger} \, \tau_{I}  \, D_{\mu} \phi)  \,   (\bar{q}^i_L \, \gamma^{\mu} \, \tau_{I}  \,  q^j_L)  \, ,
\label{eq:op5}
\end{equation}
\\[-0.3cm]
\begin{equation}
{\cal O}^{ij}_{u \phi} \, =   (\phi^{\dagger}  \phi) \, (\bar{q}^i_L \,  u^j_R \, \tilde{\phi})  \, , \quad
\label{eq:op6}
\end{equation}
where $B^{\mu \nu}$ and $W^{I}_{\mu\nu}$ are the $U(1)_Y$ and $SU(2)_L$ field 
tensors, respectively. 
As was shown in~\cite{AguilarSaavedra:2008zc}, for all the operators in~(\ref{eq:op4}) and 
(\ref{eq:op5}),  ${\cal O}^{ij}$ and ${\cal O}^{ji}$ are not independent. In fact,
by writing the combinations ${\cal O}^{i+j}$ and ${\cal O}^{i-j}$ and using
the equations of motions, it can be shown that only one of these combinations
is independent.
 This means that the number of independent operators in~(\ref{eq:op4}) and 
(\ref{eq:op5}) is reduced to three (for each light flavour). 
The above discussion leads us to the conclusion that the minimal number of 
non-4F operators needed to study top FCNC physics is 9 for each light flavour ($u$ and $c$).

The equations of motion used to reduce the number of operators relate the
operators from the strong and electroweak sectors with the 4F operators.
The number of independent 4F operators depends on the process 
considered. Following~\cite{AguilarSaavedra:2010zi} we found the minimal
number of independent 4F operators needed for t plus quark production, 
$pp \to t q$, which are shown in appendix B. 
The final lagrangian for the study of single top production via FCNC currents can then be written as
\begin{eqnarray}
{\cal L}_{qq,qg,gg \rightarrow t\, {\bar q}}&=& \frac{1}{\Lambda^2} \, \sum_{\substack{i,j=1,3 \\ or \\ i,j=2,3 \\ i\neq j }} \Big( 
\alpha_{uG \phi }^{ij} {\cal O}_{u G \phi}^{ij}
+\alpha_{uW \phi }^{ij} {\cal O}_{uW \phi}^{ij}
+\alpha_{u B \phi}^{ij}\, {\cal O}_{u B \phi}^{ij} 
+\alpha_{\phi u}^{ij}\, {\cal O}_{\phi u}^{ij}+\alpha_{\phi q}^{(3,ij)}\, {\cal O}_{\phi q}^{(3,ij)}  
\nonumber\\
&& 
+\alpha_{\phi q}^{(1,ij)} \,  {\cal O}_{\phi q}^{(1,ij)} 
+\alpha_{u \phi} \, {\cal O}_{u\phi}^{ij}  \Big) +\frac{1}{\Lambda^2}  \, {\cal L}_{4fu}+
\frac{1}{\Lambda^2} \, {\cal L}_{4fc}
\end{eqnarray}
where ${\cal L}_{4fu}$ and ${\cal L}_{4fc}$ are described in appendix B.
One should note that in order to keep a manageable number of 4F
operators we only consider initial states with up-quarks in the hadron
colliders case. There are reasons for considering a reduced set of 4F
operators, namely the ones that have in the initial state either $uu$
and $u \bar u$. First, and assuming that all 4F coupling constants are of the same order,
these initial states provide the largest contribution
for the cross section. Second, our main goal
is to provide a means to distinguish between operators by analysing 
different distributions and this can only be done if the number of operators
is not too large. The addition of further 4F operators will
be done in the future if found necessary.
The minimal number of 4F operators in the case 
of FCNC $pp \to t \bar t$ was recently considered in~\cite{Biswal:2012mr}.

Before ending this section we will briefly discuss the bounds on
the coupling constants $\alpha_i$. In the effective operator approach, the lagrangian has the SM symmetries.
Therefore, physics of the top quark is related with B physics. In 
reference~\cite{Fox:2007in}  this relation was explored in
order to constrain the electroweak FCNC operators~\footnote{Other analysis based on B physics observables and 
electroweak precision constraints were also performed
in~\cite{Drobnak:2011wj} leading to similar conclusions.}. 
The most constrained operators are obviously the ones 
built with quark doublets only while the less constrained
are the ones built with quark singlets only.   
Consequently, for the first generation, bounds on operators with doublets only,
$\alpha_{\phi q}^{(3,ij)}/\Lambda^2$ and $\alpha_{\phi q}^{(1,ij)}/\Lambda^2$,
are of the order $0.01$ TeV$^{-2}$. For 
operators $\alpha_{uW \phi}^{ut}/\Lambda^2$ and $\alpha_{u B \phi}^{ut}/\Lambda^2$
the bounds are of the order $0.3$ TeV$^{-2}$ while for 
$\alpha_{uW \phi}^{tu}/\Lambda^2$ and $\alpha_{u B \phi}^{tu}/\Lambda^2$
we have $1$ TeV$^{-2}$. Finally, regarding operators 
with singlets only, like $\alpha_{u \phi}/\Lambda^2$,  reference~\cite{Fox:2007in}  
obtained a bound of the order $3$ TeV$^{-2}$. The bounds
for operators relating the second and third generation
are of the same order of magnitude.

As stated in the introduction, there are new direct bounds from the LHC
that lowered the limit on $BR (t \to q Z)$ 
to $0.34 \%$~\cite{CMStqZ}.  A new indirect
bound from HERA~\cite{Abramowicz:2011tv} is also available $BR (t \to q \gamma) < 0.5 \%$.
 Also, a combined study on B physics and Tevatron data on top 
quark production cross section places an indirect bound on the sum of the 
FCNC branching ratios forcing them to be below the percent level~\cite{Ferreira:2009bf}.
All these new bounds do not imply any dramatic changes on the bounds
in the electroweak sector.

Regarding the strong FCNC operators the most recent
search is the one from the ATLAS collaboration~\cite{Aad:2012gd} in direct top production at the LHC.
The upper limits obtained at 95 \% CL for the strong couplings
are  $\kappa_u/\Lambda < 6.9 \times 10^{-3}$ TeV$^{-1}$ and $\kappa_c/\Lambda < 1.6 \times 10^{-2}$ TeV$^{-1}$
which in turn can be translated into strong branching ratio bounds
$BR (t \to u g) < 5.7 \times 10^{-5}$ and $BR (t \to c g) < 2.7 \times 10^{-4}$.

Contrary to the the dimension six FCNC operators from
the strong and electroweak sector, there are no useful
bounds on the four fermion operators involving two top quarks and
this is even more so if the top is right-handed. Therefore,
the LHC can place constraints on these operators.

\section{Physical processes}

MEtop generates events according to the Von Neumann algorithm (see~\cite{PDG} for details).
The amplitudes for each process were generated with CalcHEP~\cite{Pukhov:1999gg}, 
and the Feynman rules for the effective operators were derived with 
LanHEP~\cite{Semenov:1998eb}. 
Integrations are performed using the CUBA
library~\cite{Hahn:2004fe}, configured to use VEGAS algorithm~\cite{Vegas}
Generation of events for hadron colliders need the linking with 
the LHAPDF package~\cite{Whalley:2005nh}. The events are written
in the standard format of Les Houches event file~\cite{Alwall:2006yp}  (.LHE).
Whenever possible our results were checked via a completely different path.
First, the Feynman rules were generated by the implementation of the 
effective operators in UFO~\cite{Degrande:2011ua}. Then, cross sections calculation and event generation
was performed using MadGraph 5~\cite{Alwall:2011uj}. We always found an excellent agreement
with MEtop.

The following hard processes are already included in MEtop: direct top 
production at LO and NLO both for $pp$ and for $p \bar p$ colliders, which at the parton level
amounts to the processes $g q \to t$ and
$g u \to g t$ where $q=u,c$; $pp (\bar p) \to q t$, with all possible parton
contributions in the initial state taken into account
and $q$ is now any quark other then the top-quark. The corresponding 
conjugate processes, with an anti-top in the final state, are also included, and 
the processes can be generated independently. 
There are several available packages in MEtop, each containing
a different set of operators. The list of operators in each
package is presented in appendix C.

In figure~\ref{fig:Dtop} we present direct top production together
with the top+gluon processes. The two have in common the fact 
that only the strong FCNC couplings contribute to the process. 
Only one diagram contributes to direct top channel while several
diagrams with gluon exchange contribute to top+gluon.
MEtop allows for generation of events at LO and at NLO. 
As we have already mentioned, currently, FCNC direct top production
events can be generated at LO with the PROTOS generator.
\\
\begin{figure}[h!]
\begin{center}
\hspace{-1.2cm}
\includegraphics[width=4.5in,angle=0]{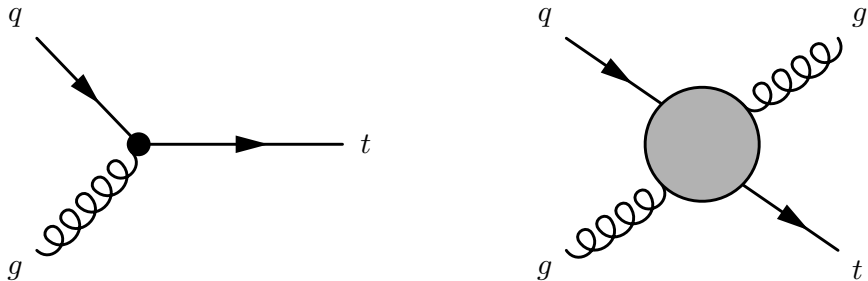}
\vspace{-0.3cm}
\caption{FCNC leading order direct top production and top + gluon production at the parton level. 
Only FCNC strong operators contribute to the process.}
\label{fig:Dtop}
\end{center}
\end{figure}

\begin{figure}[h]
\begin{center}
\hspace{-1.2cm}
\includegraphics[width=4.5in,angle=0]{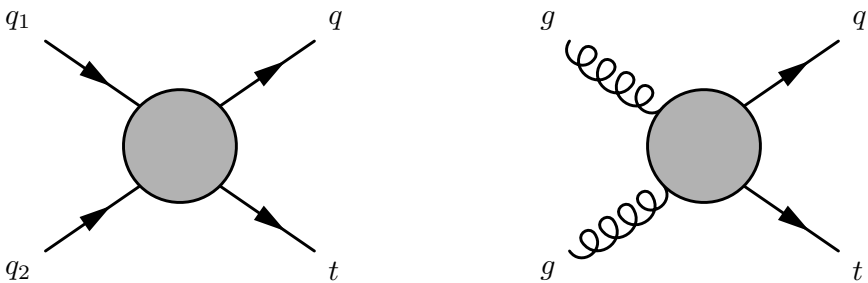}
\vspace{-0.3cm}
\caption{FCNC leading order top + quark production at the parton level. FCNC strong and 
electroweak operators contribute to the process together with 4F operators.}
\label{fig:Lqtop}
\end{center}
\end{figure}

In figure~\ref{fig:Lqtop} the diagrams for top + quark production are shown. Both  $q_1, \, q_2$ and
$q$ run through all quarks other then the top-quark, that is $u,d,c,s,b$ and 
the respective anti-particles. Again, conjugate processes are also included.
In the diagram on the right, only strong FCNC operators   
are present. On the left diagram all operators can take part, including the 4F ones, contributing
to LO single top production at the parton level.
A detailed description of each process will be presented in appendix C.

\section{NLO approximation to direct top production}

In an event generator, initial (ISR) and final (FSR) state radiation
is included trough a shower mechanism. The shower mechanism
assumes collinear factorization, that is, that the real radiation process
can be approximated by a branching mechanism, where the first
QCD radiation is emitted by one of the legs of the Born configuration. 
In equation \ref{eqn:Shower} we present the relation between the transition
amplitude for the case of $q \to qg$ splitting
\begin{equation}
|M_{n+1}|^2 d\Phi_{n+1} \implies |M_{n}|^2 d\Phi_{n} \frac{\alpha_S}{2\pi} \frac{dt}{t} P_{q,qg} (z) dz \frac{d\phi}{2\pi}
\label{eqn:Shower}
\end{equation}
where $M_{i}$ is the amplitude and $d\Phi_{i}$ is the phase space
for the $i$th body processes and $P_{q,qg} (z) $ is the 
Altarelli-Parisi splitting function. This approximation breaks down
in the hard $P_T$ region where the matrix element $M_{n+1}$
should be used.  A factorization prescription or matching
scheme is then used to merge these two regions in a smooth
and optimised way (see for example the merging approaches 
CKKW~\cite{Catani:2001cc} and MLM~\cite{MLM}).  

In the previous section we have presented in figure~\ref{fig:Dtop} 
the parton level contributions to direct top production, together
with the parton level contribution to the hard process $g q \to g t $.
The later process contributes
to the inclusive direct top production. The $g q \to g t $ process has
soft and collinear divergences, and this problem can only be solved by including the NLO corrections.
Furthermore, the FCNC direct top production cross section was
calculated in~\cite{Liu:2005dp} and a considerable enhancement of about 40 \%
was found relative to the leading order cross section for the LHC with $\sqrt{s} = 14$ TeV. 
Therefore it is desirable to have a NLO generator for
direct top production at the LHC. It was also shown in~\cite{Liu:2005dp}
that the NLO QCD corrections vastly reduces the dependence of the total cross
section in the renormalization and factorization scales which in turn
increases the confidence in the predictions.

\begin{figure}[h]
\begin{center}
\hspace{-1.2cm}
\includegraphics[width=4.5in,angle=0]{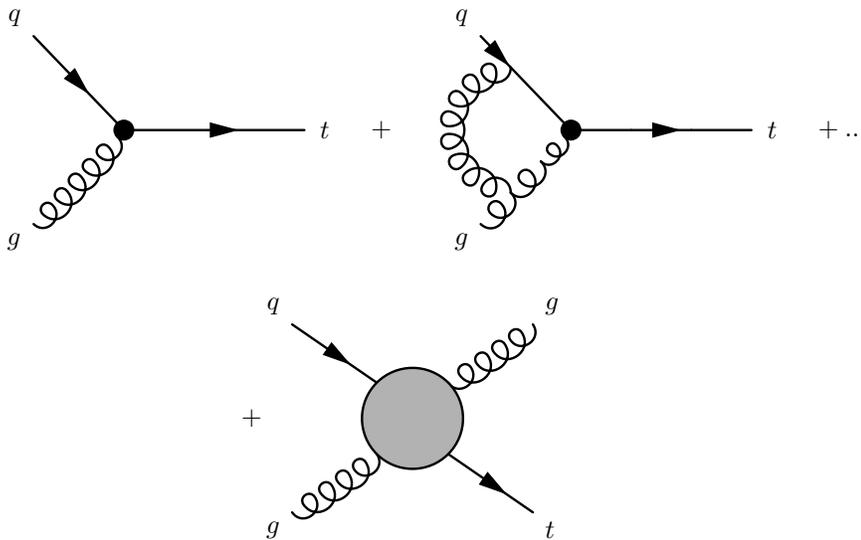}
\vspace{-0.3cm}
\caption{Inclusive FCNC direct top production at NLO in QCD.}
\label{fig:DtopNLO}
\end{center}
\end{figure}

\vspace{1cm}

In order to calculate the inclusive NLO FCNC direct
top cross section one has to consider the Born diagram,
the virtual contributions and the real emission diagrams.
A sketch of the diagrams from each of the above mentioned
contribution are shown in figure~\ref{fig:DtopNLO}.
It is well known that at NLO, the integration in the full phase-space  
gives rise to infrared divergence from the virtual-born interference part
and these divergences are cured by the addition of the real emission process. 
Although easily fixed for a total cross section analytical calculation, this problem is not 
straightforward to deal with at an event generator level. In fact,
there is no way to extract the infrared infinities as it is usually done
with dimensional regularization in a theoretical calculation. There are methods to deal
with these infinities such as Phase Space Slicing~\cite{Baer:1989, Giele:1992, Giele:1993} and
Subtraction Methods~\cite{Catani:1997a, Catani:2002a}. 
In this work we will adopt an effective NLO approximation~\cite{Boos:2006af} to simulate direct top events at the NLO level. 
In this approach  a merging  scheme between $2 \to 1 $ and $2 \to 2$ events is performed, where each process
 will separately populate two  distinct but joint regions of the phase-space. A resolution parameter must then be defined, 
 which in the present case is the cut in transverse momentum of the top quark applied to the real radiation process.  
 This $P_T^{cut}$ will then play  a role of a matching variable, $P_T^{match}$. The phase-space region for small $P_T$ will be described by 
 the $2 \to 1 $ process and the subsequent parton shower (PS) mechanism, whereas the hard $P_T$ region will be described by 
 the $2 \to 2$  process. One must then just make sure that the transition is done in a smooth way. The virtual corrections are 
 included via a K-factor applied to the cross section of the $2 \to 1 $ process. We assume this to be a good approximation because
  the kinematics of the Born and Virtual configurations of the direct top process should be very similar. 
  The events will then be generated according to the following relation
\begin{equation}
\sigma_{NLO} = K \sigma_{LO} (P_T^{PS} < P_T^{match}) + \sigma_{Real} (P_T > P_T^{match}) 
\label{eqn:sigmaNLO1}
\end{equation}
where $\sigma_{LO}$ is the tree-level direct top contribution, $ \sigma_{Real}$ is the real radiation part, 
K is the  K-factor and $P_T^{PS}$ and $P_T^{match}$ are the transverse momentum of the first
 PS emission and the integration cut of the real radiation process, respectively. Once the direct
top events are produced, they will be radiated through a radiator
like the one in PYTHIA \cite{PYTHIA6}. In order to avoid double 
counting, the matching must ensure that the first PS emission from the $2 \rightarrow 1$
 process will not fall within the  $2 \rightarrow 2$ 
configuration phase-space. There are two ways of accomplishing it: either by vetoing 
all radiated $2 \rightarrow 1$  events  that would be within the $2 \rightarrow 2$
configuration phase-space or simply by limiting
the phase-space region of the radiated $2 \rightarrow 1$ events to the boundaries
defined by the resolution variable. We choose to adopt the later.

In order to follow this approach, one must ensure that the PS  
mechanism added to the generated events from the Born configuration will populate the region with 
$P_T < P_T^{match}$, which can be assured using a PT-ordered shower~\cite{Sjostrand:2004ef},  available in both 
current PYTHIA versions 6.4 and 8.1. We therefore assume
 that the generated events will be showered by a PT-ordered mechanism.
Therefore we start by calculating the three cross sections 
from equation  (\ref{eqn:sigmaNLO1}), with $P_T^{cut} = P_T^{match}$ for the $2 \to 2$ process. 
For the $\sigma_{NLO}$ cross section we have used the expressions from~\cite{Liu:2005dp}, where
the top quark is on-shell. The tree-level direct top and top+gluon amplitudes were 
generated with CalcHEP where the top quark and the W decays were included in order to preserve
spin correlations. The cross sections are then calculated with the Cuba library.  Hence,
the K factor is calculated "on the fly" for each sub-process.
After extracting the K-factor, the events are generated weighed according to equation~\ref{eqn:sigmaNLO1}. 
The PS starting scale can then be configured to start the branching in $P_T^{match}$ for the $2 \to 1$ 
events configuration, which 
in MEtop is done by preparing the .lhe files to be used by PYTHIA. 
%
%
A short remark is in order - in the $2 \to 1$ configuration, no meaningful FSR from the
top quark can be present due to its large mass. Hence, we consider a 
good approximation to take only ISR into account.

\begin{figure}[h]
\begin{center}
\includegraphics[scale = 0.6]{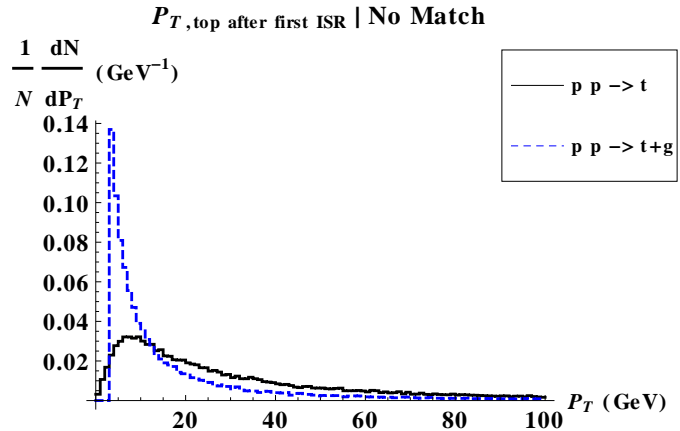}
\caption{$P_T$ distribution of the top quark for $\sqrt{s}$ = 7 TeV. The black solid line is for direct top production after the first branching in ISR,
 with starting scale of $m_{t}$. The blue dashed line is for the hard process top+gluon production.}
\label{fig:NoMatch}
\end{center}
\end{figure}
In figure~\ref{fig:NoMatch}, the black solid line represents the $P_T$ distribution of the top quark
in direct top production,  after the first branching in ISR,
 with starting scale of $m_{t}$. In the same figure, the blue dashed line
 represents the hard process: top+gluon production. As described
 previously, $P_T$ is the kinematical variable chosen to match the two
 processes avoiding double counting in the low $P_T$ region.

\begin{figure}[h!]
\centering
\hspace{-1.cm}\includegraphics[width=3.0in,angle=0]{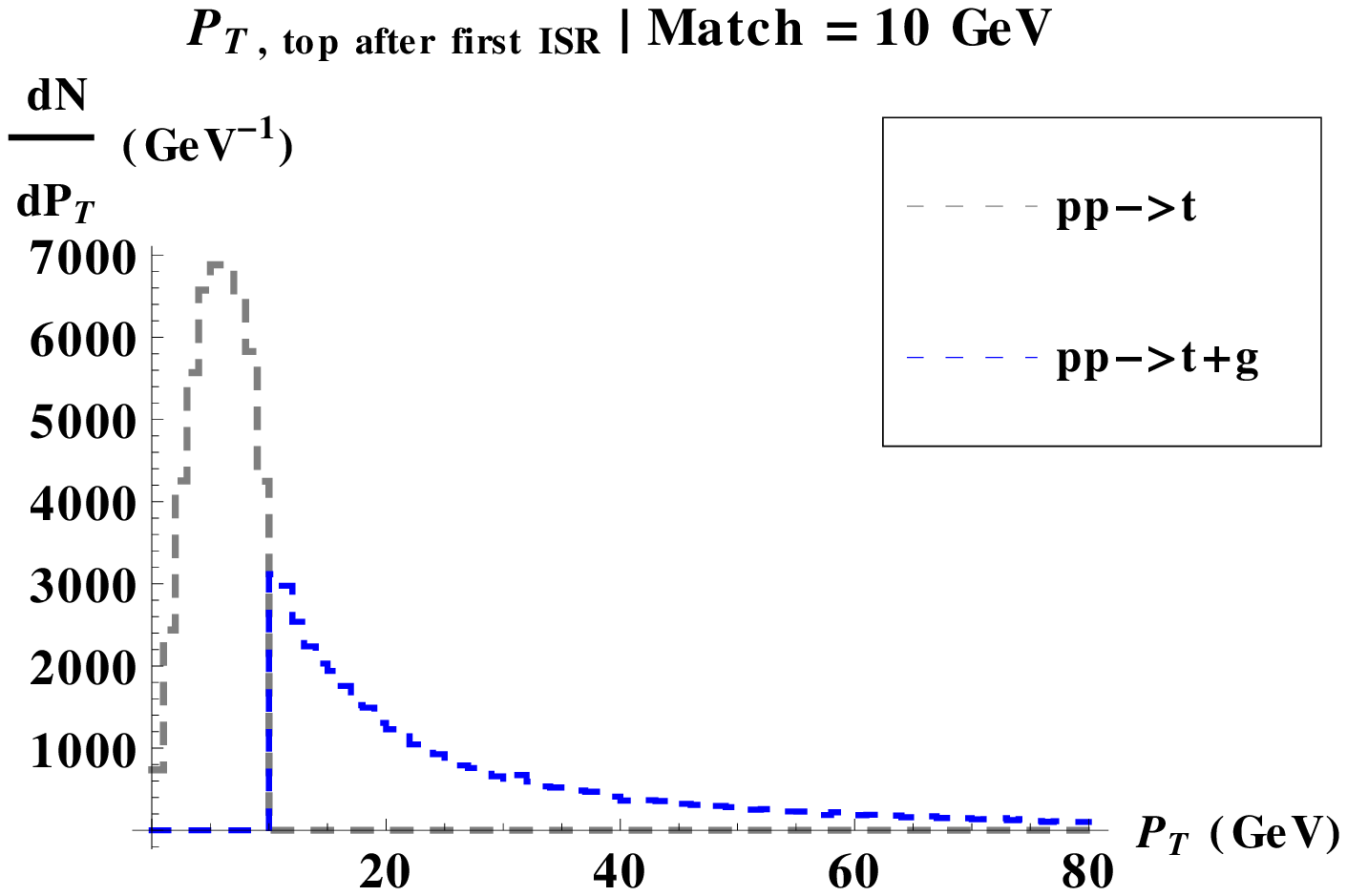}
\hspace{-.3cm}
\includegraphics[width=3.0in,angle=0]{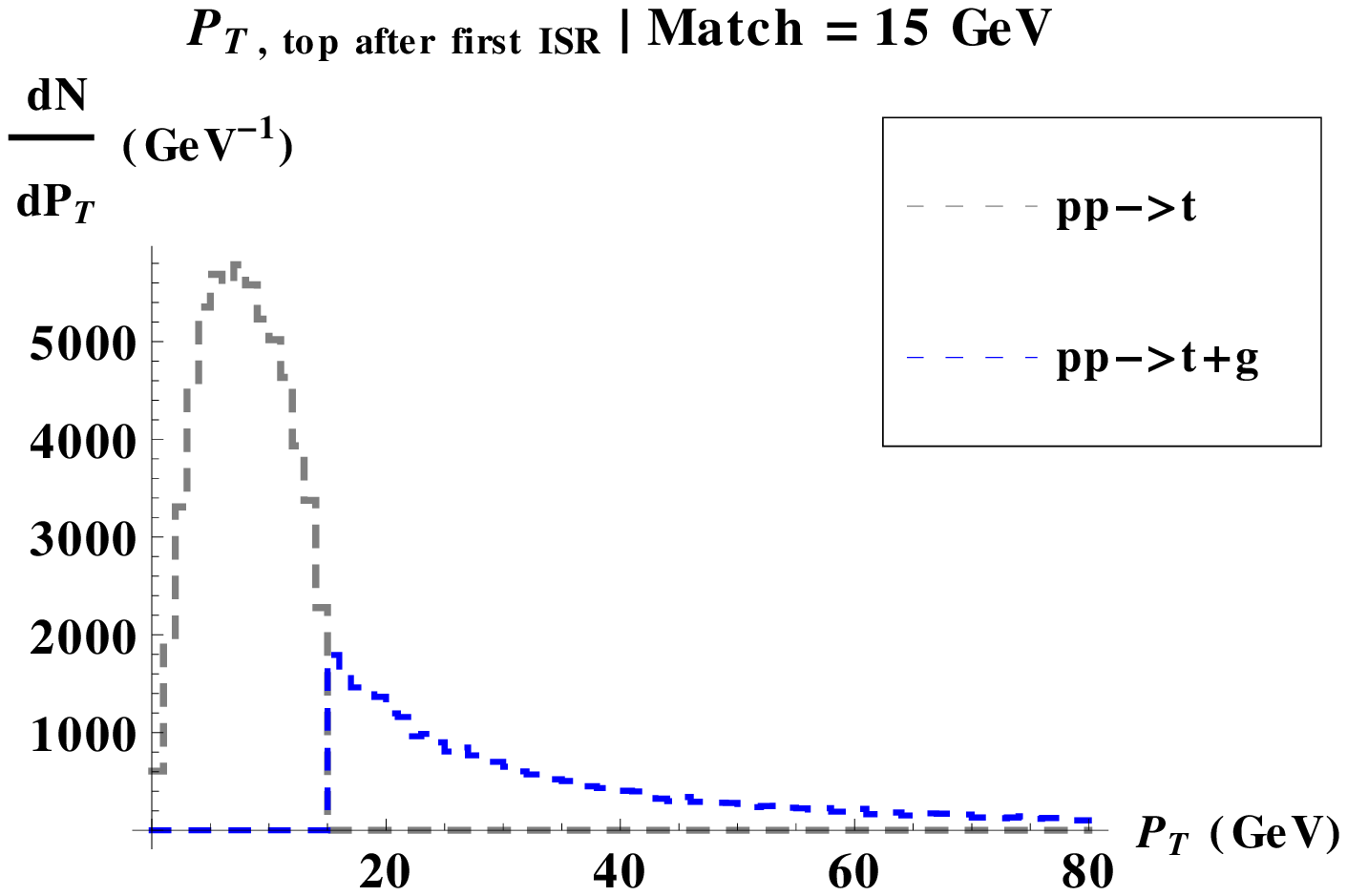}
\caption{$P_T$ distribution of top quark after the first ISR branching with a $P_T^{match} $ of 10 GeV (left)  and 15 GeV (right). }
\label{fig:ptmatch510}
\end{figure}
In figure~\ref{fig:ptmatch510} we present the $P_T$ distribution of the
top quark after the first ISR branching
with a $P_T^{match} $ of 10 GeV (left)  and 15 GeV (right).  
The natural criterion to determine the value of the $P_T$ 
matching parameter in the effective NLO approximation
is  the smoothness of the transverse momentum distribution. There are no significant differences when the value
of $P_T^{match} $ is varied in the 5 GeV to 20 GeV range. As can be seen in figure~\ref{fig:ptmatch510} , 
there is never a completely smooth transition between the two sets of events. This effect 
should be included as part of the systematic uncertainties. This feature
was checked for a large range of $P_T$ match. 
After including the full shower (ISR+FSR) and Multiple Interaction (MI) we have
opted for a value of $P_T^{match} $ of 10 GeV.

\begin{figure}[h!]
\centering
\hspace{-1.cm}\includegraphics[width=3.0in,angle=0]{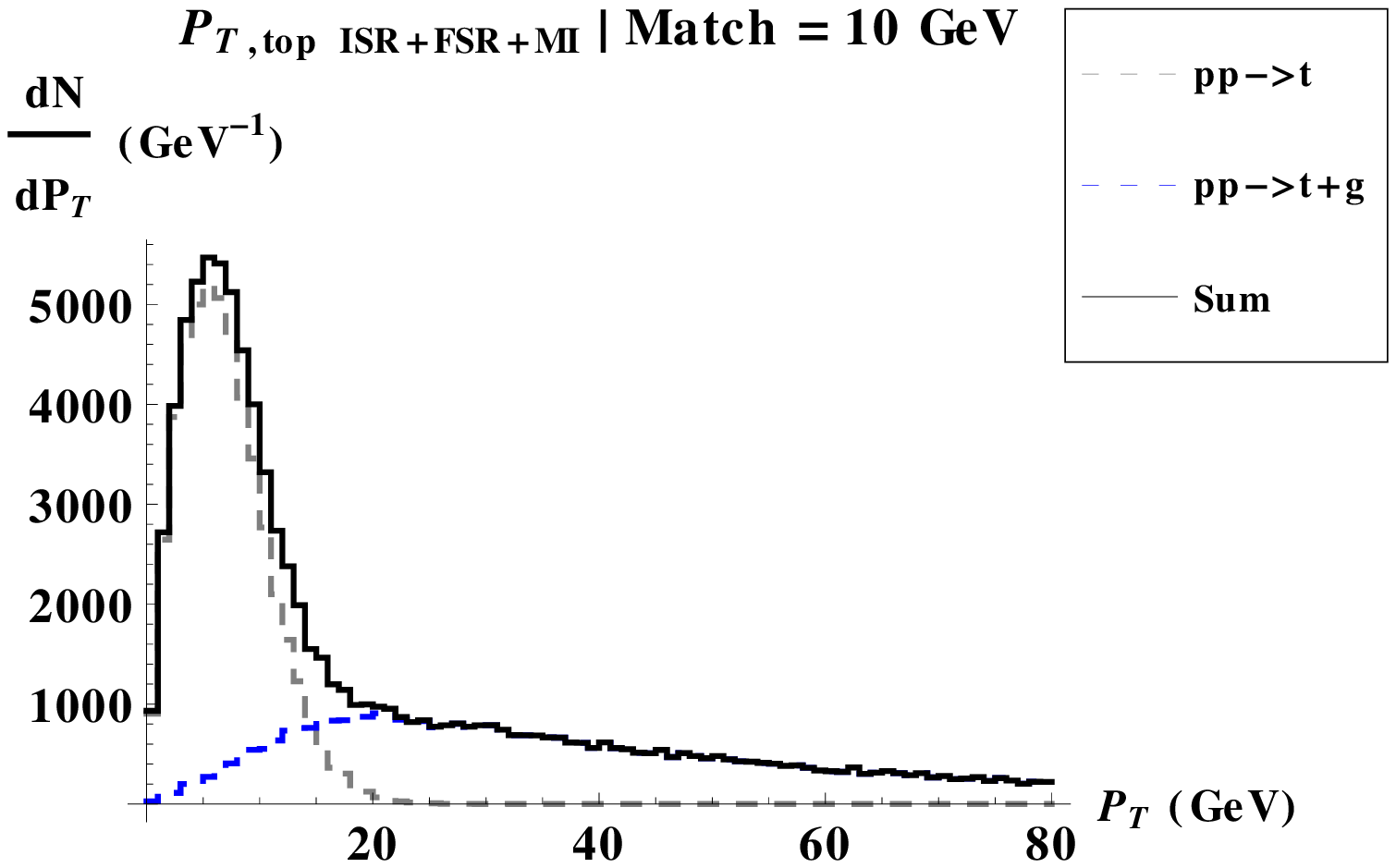}
\hspace{-.3cm}
\includegraphics[width=3.0in,angle=0]{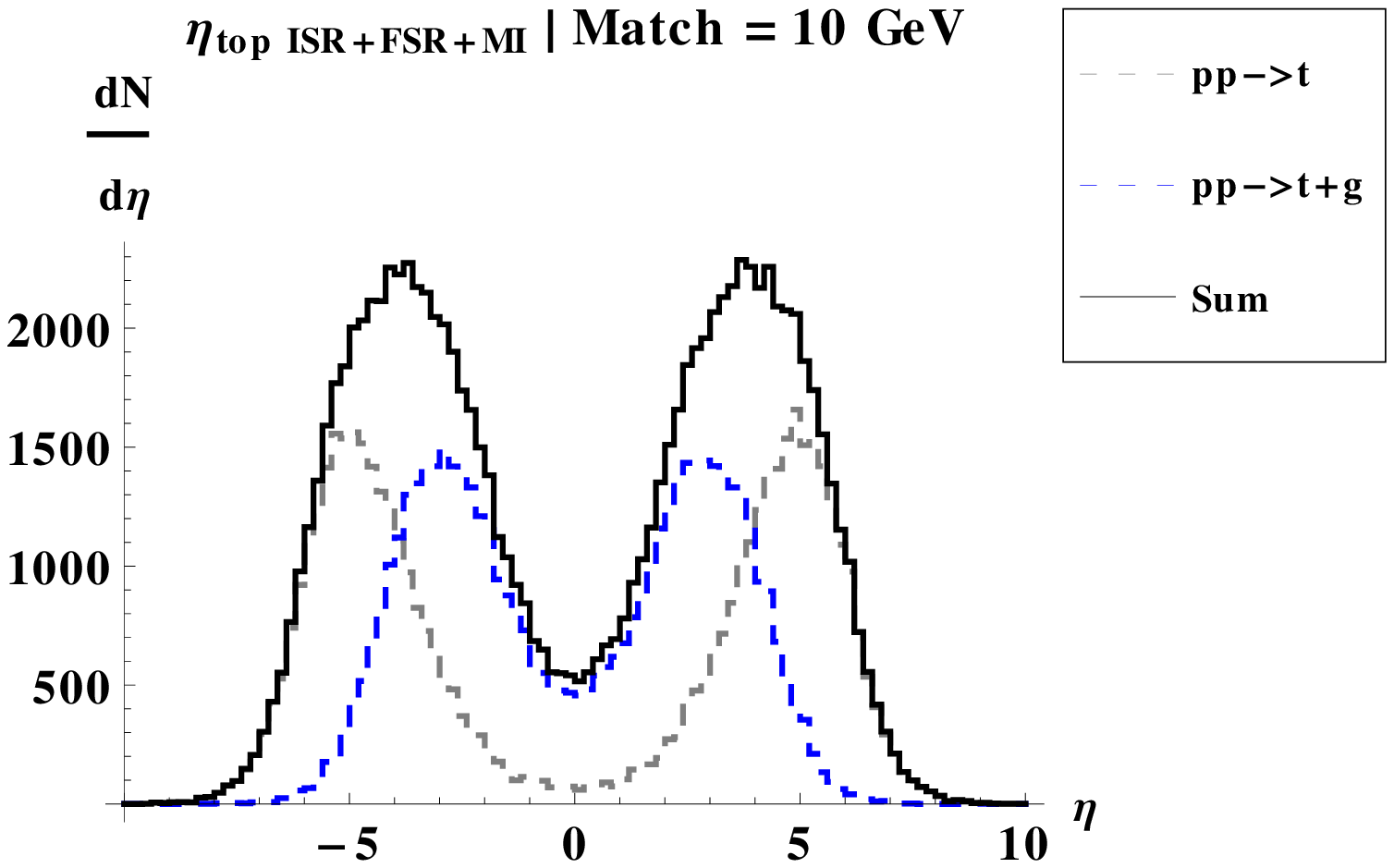}
\caption{$P_T$ (left)  and $\eta$ (right) distributions of the top quark at the partonic level after the full shower (ISR+FSR) and Multiple Interaction. }
\label{fig:ptandetamatch10}
\end{figure}
In figure~\ref{fig:ptandetamatch10} we show the $P_T$ (left)  and $\eta$ (right)
 distributions of the top quark at the partonic level after the full shower and MI
 for $ P_T^{match} = 10$ GeV. The blue dashed line represents the real
 radiation part while the grey dashed line is the direct top fully showered but
 with the $p_T$ starting scale at 10 GeV. The solid black line is the final NLO 
 distribution which amounts to the sum of the previous two.

\begin{figure}[h!]
\centering
\hspace{-1.cm}\includegraphics[width=3.0in,angle=0]{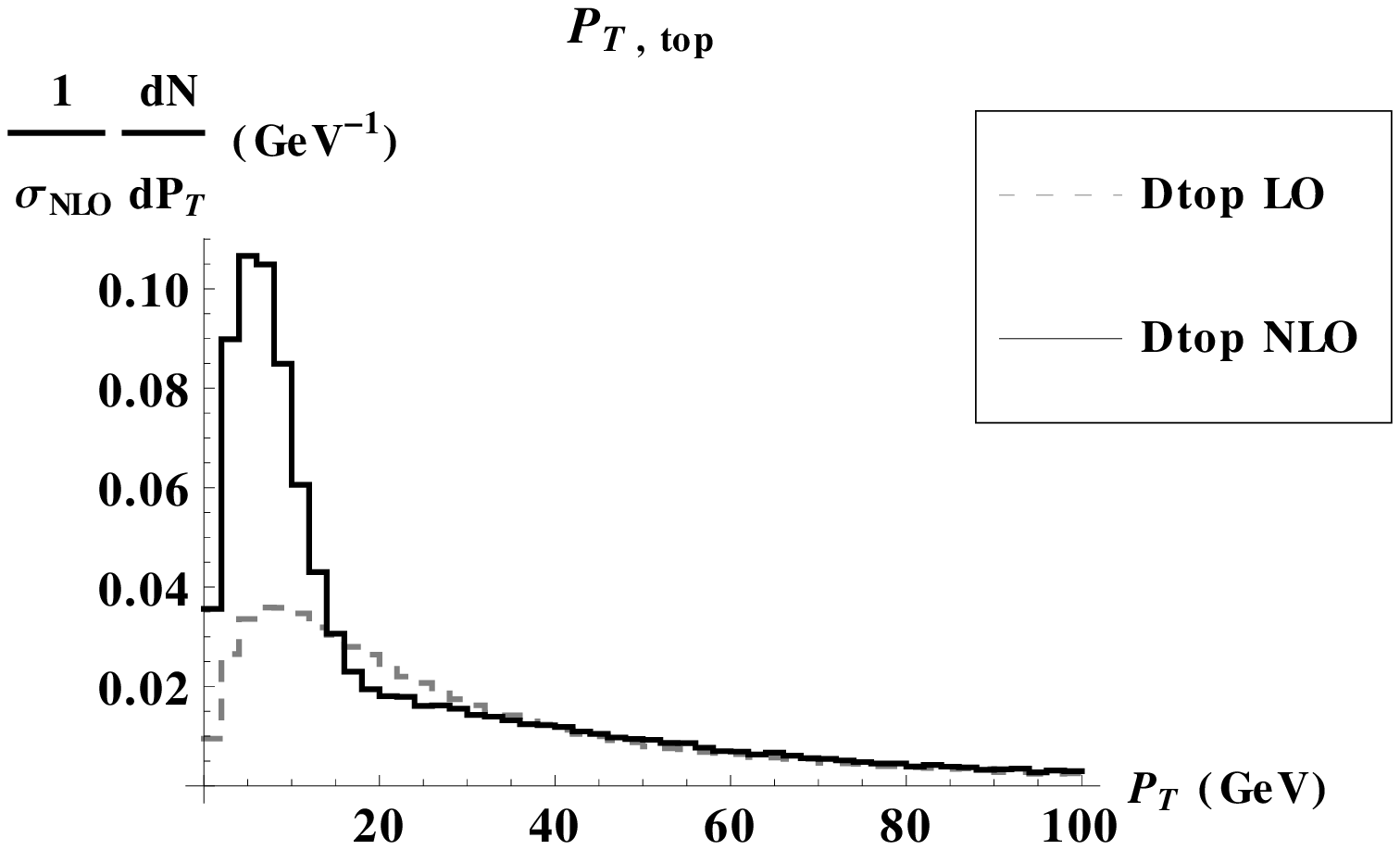}
\hspace{-.3cm}
\includegraphics[width=3.0in,angle=0]{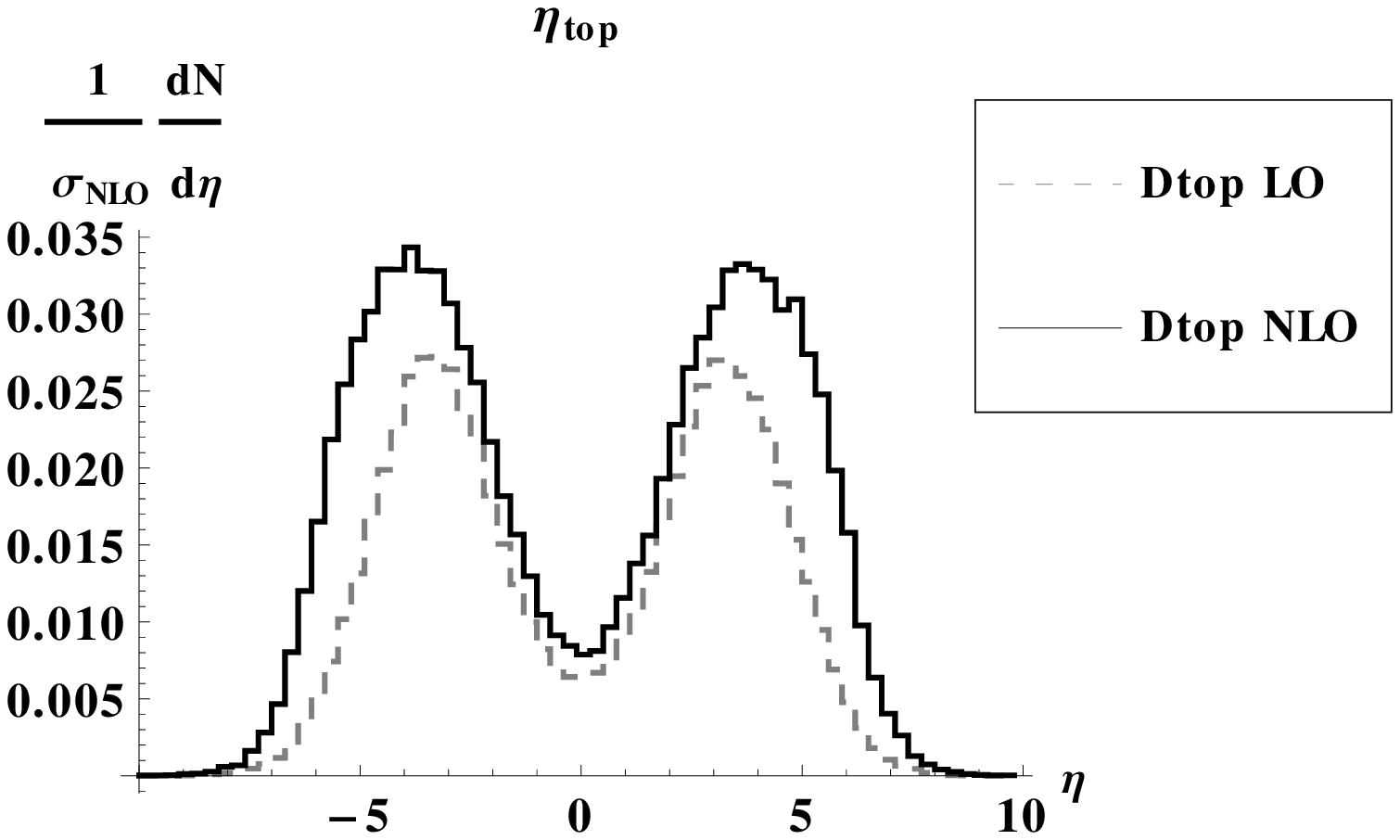}
\caption{Comparison of the LO and NLO $P_T$ (left)  and $\eta$ (right) distributions of the top quark at the partonic level after the full shower (ISR+FSR) and Multiple Interaction. }
\label{fig:LOvsNLOtop}
\end{figure}  
In figure~\ref{fig:LOvsNLOtop} we present the LO and NLO $P_T$ (left)  
 and $\eta$ (right) distributions of the top quark at the partonic level after the full shower and MI.
 It is clear from figure~\ref{fig:LOvsNLOtop} that
 the $P_T$ and $\eta$ distributions  of LO direct top production are
 quite different from the corresponding NLO direct top ones. 
 In fact, the distributions
 show that the use of a constant K factor does not correctly describe the behaviour 
 of direct top at NLO. Hence, a new analysis is needed to improve the accuracy of
 the bounds on the strong coupling constants $\kappa_u$ and $\kappa_c$.  
The direct top NLO $p_T$ distribution is shifted to low values of $p_T$ as compared
to the LO distribution while the $\eta$ distributions are shifted to higher values
of $\eta$ as compared to the LO one.

\begin{figure}[h!]
\centering
\hspace{-1.cm}\includegraphics[width=3.0in,angle=0]{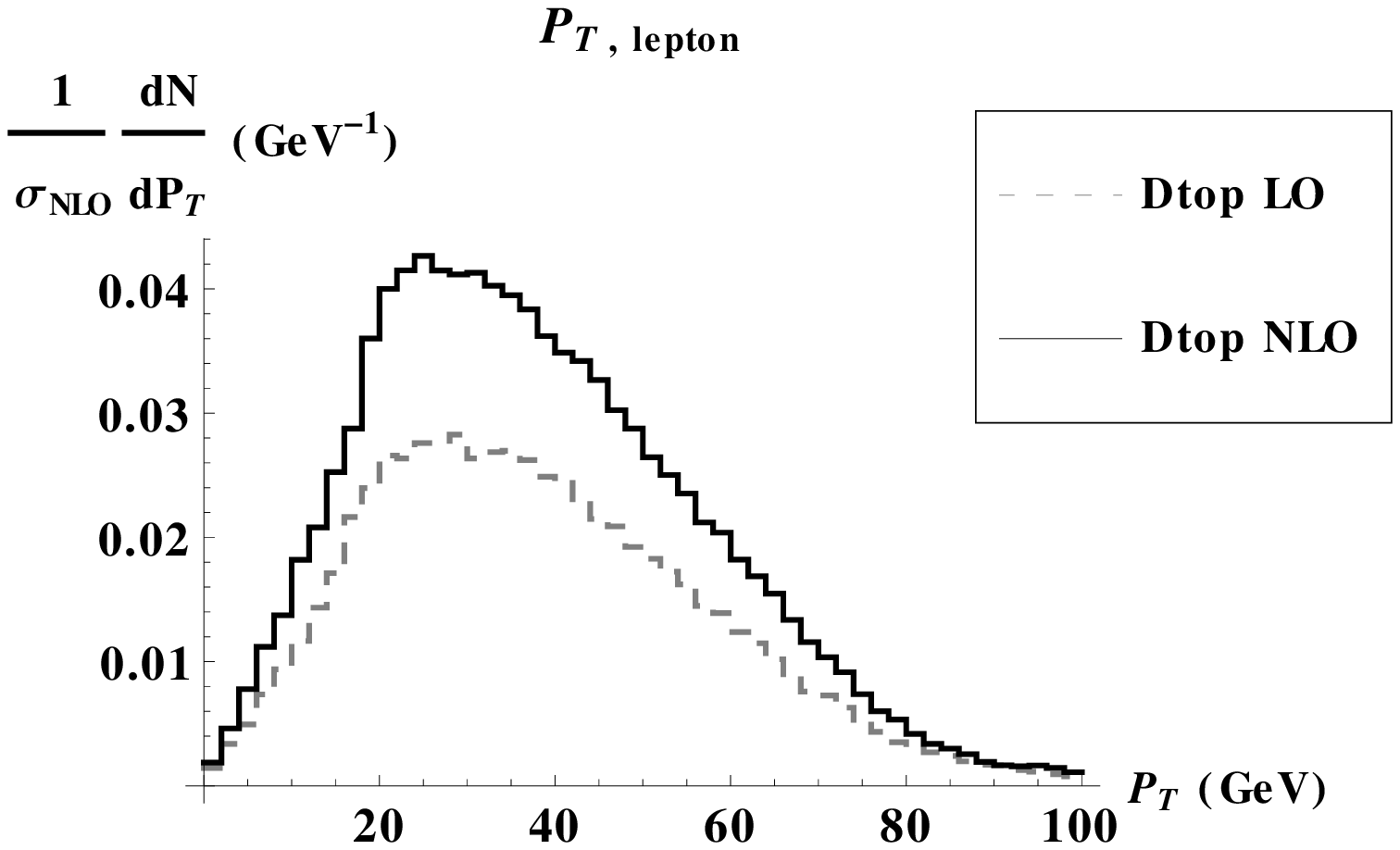}
\hspace{-.3cm}
\includegraphics[width=3.0in,angle=0]{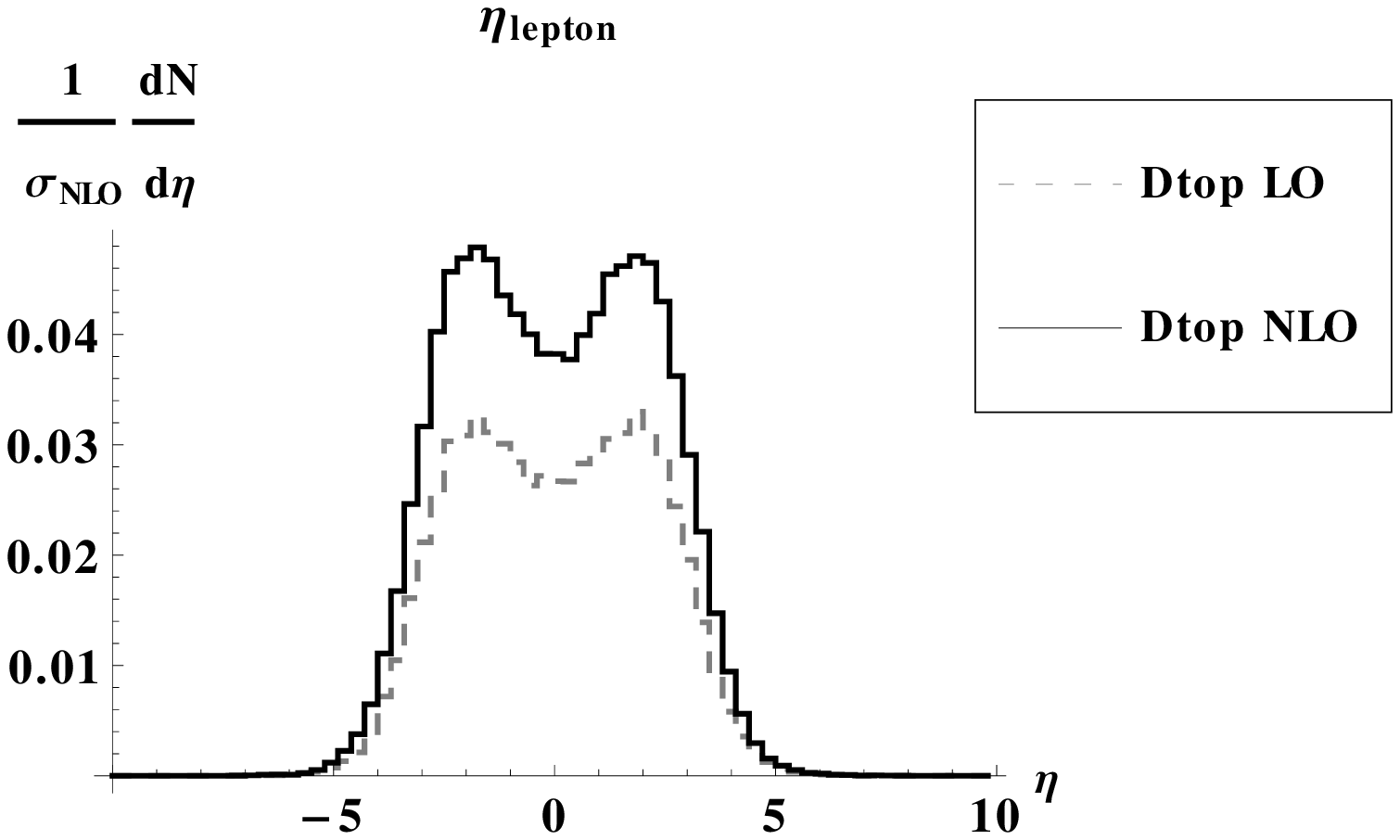}
\caption{Comparison of the LO and NLO $P_T$ (left)  and $\eta$ (right) distributions of the lepton from $t \to bW \to b l \nu$ at the partonic level after the full shower (ISR+FSR) and Multiple Interaction. }
\label{fig:LOvsNLOlepton}
\end{figure}

The actual experimental analysis is performed by looking at the distributions
of the final state particles. Therefore, in figure~\ref{fig:LOvsNLOlepton} we present
the comparison between LO and NLO $P_T$ (left)  and $\eta$ (right) distributions 
of the lepton from $t \to bW \to b l \nu$ at the partonic level after the full shower and MI.
Again, it is clear that the level of improvement by considering the NLO distributions
heavily depends on the particular analysis being performed.

\begin{figure}[h!]
\centering
\hspace{-1.cm}\includegraphics[width=3.0in,angle=0]{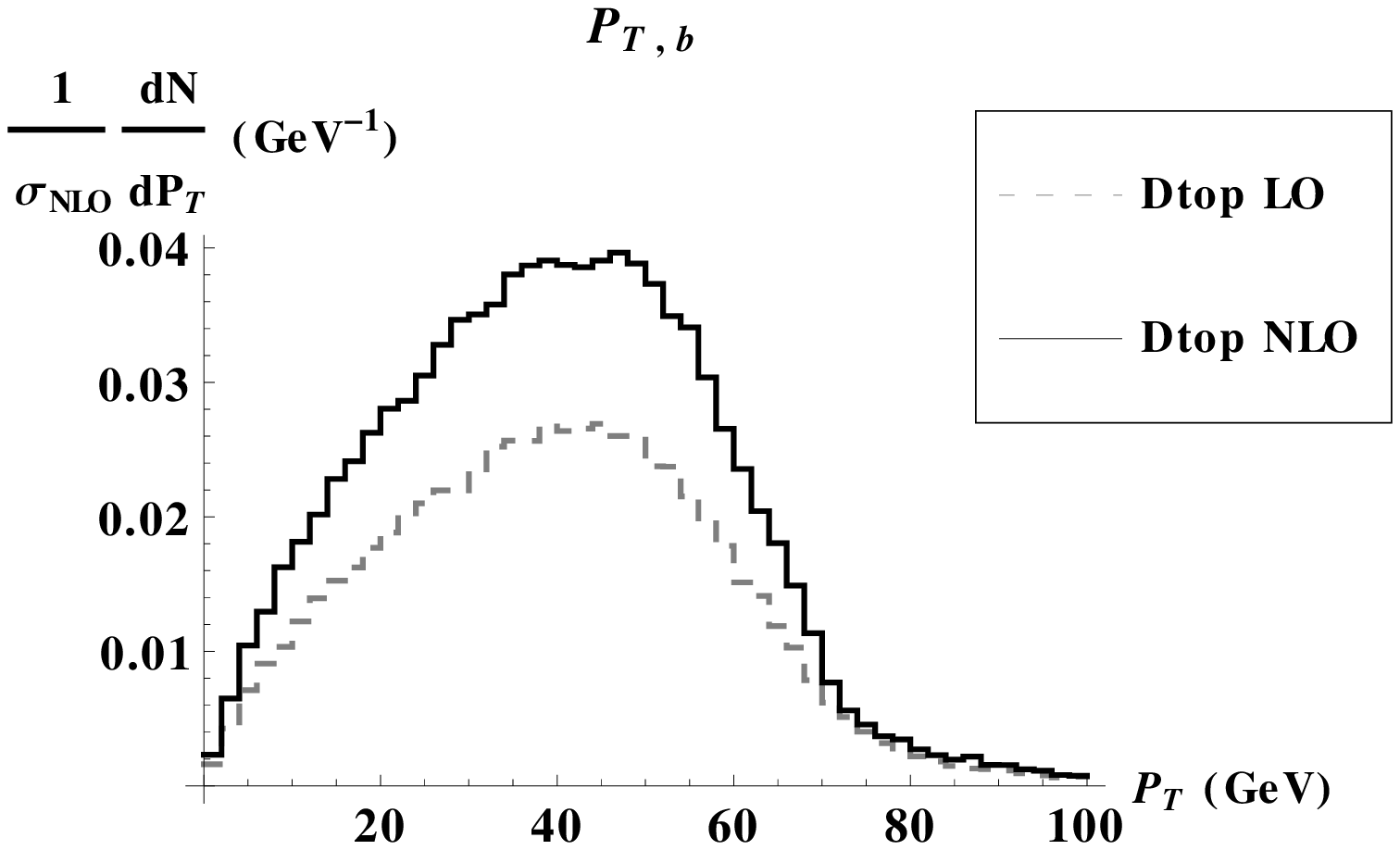}
\hspace{-.3cm}
\includegraphics[width=3.0in,angle=0]{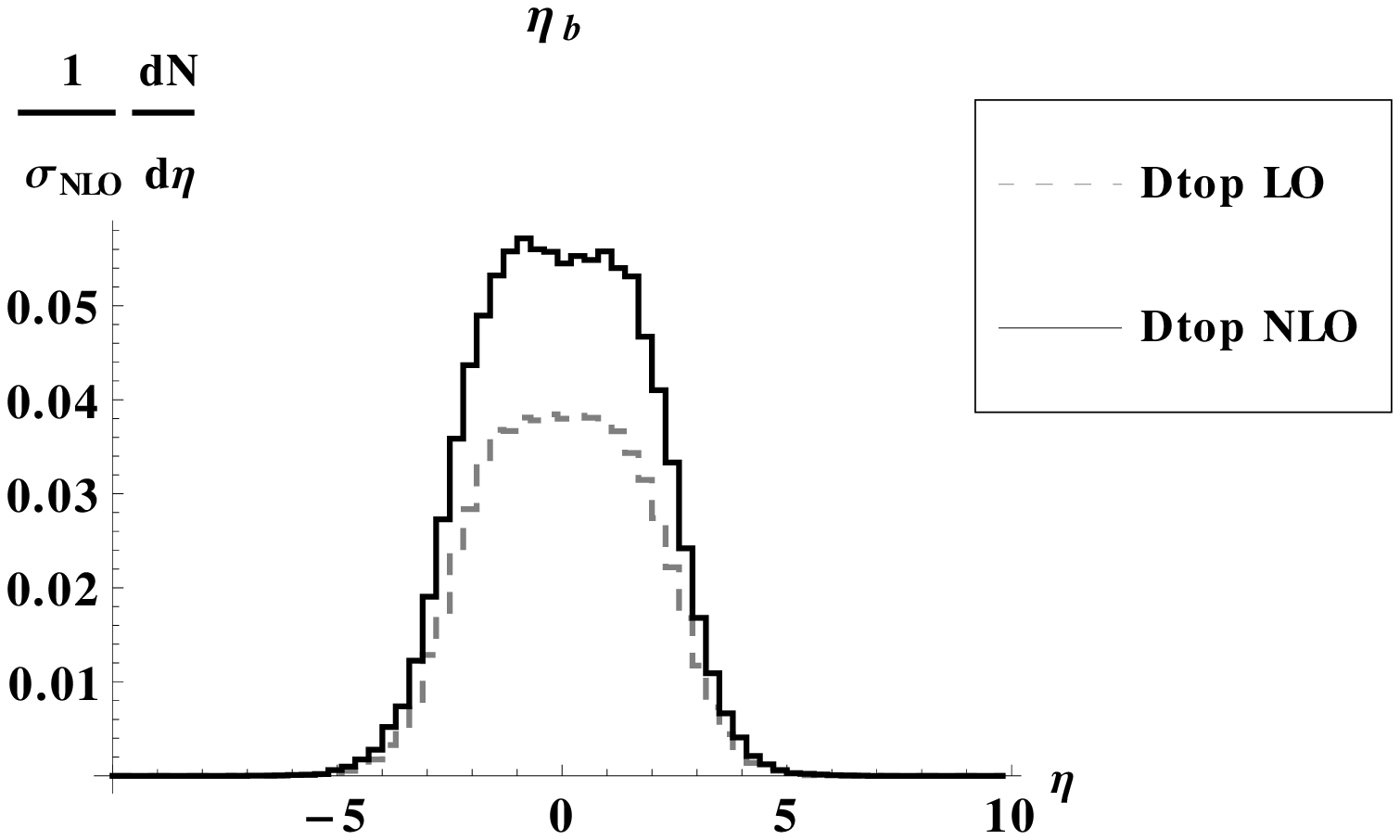}
\caption{Comparison of the LO and NLO $P_T$ (left)  and $\eta$ (right) distributions of the b-quark from $t \to bW \to b l \nu$ at the partonic level after the full shower (ISR+FSR) and Multiple Interaction.  }
\label{fig:LOvsNLOb}
\end{figure}

Finally, in figure~\ref{fig:LOvsNLOb} we compare the
 LO and NLO $P_T$ (left)  and $\eta$ (right) distributions 
of the b-quark coming from $t \to bW \to b l \nu$ at the partonic level after the full shower and MI. 

We have just described how we generate a sample of inclusive 
direct top production at NLO. However, if the goal is to
set  a limit on the strong FCNC coupling, one needs to add the
events generated in the process $pp \to t + jet$ 
composed by the parton level processes $gg \to t + q$ and 
$q_1 q_2 \to t + q$, where $q, \, q_1, \, q_2$ are quarks
other then the top and hermitian conjugate reactions are included.
Although the main contribution to top + jet production
comes from the reaction where the jet is a gluon, all
processes where the strong FCNC operator intervenes
should be taken into account in all analysis. As these 
processes also suffer from infrared and collinear  divergences, we have
decided to avoid them by using a similar cut to that
of $P_T^{match} $, that is, $p_T > 10$ GeV.
%
%
The complete NLO QCD corrections
to the FCNC process of top+jet production were presented in~\cite{Gao:2009rf}.
The corrections can increase the cross section by 10 \% to 30 \%
at the LHC@14TeV.

When generating the top + quark subprocesses we have to decide
what is considered as signal in our analysis. $pp \to tq$ has three different
classes of subprocesses: the ones which are exclusive to the Standard Model, like
$u \bar d \to t \bar b$, the ones that are originated exclusively via FCNC interactions,
e.g. $u u \to t u$, and the ones where interference
occurs, like $u \bar b \to t \bar b$. We define as FCNC signal
the contributions from the two later classes of subprocesses. For
the pure FCNC processes this poses no problems. However, for
the interference terms this procedure leads to the inclusion
of a small portion of events that will also be counted as background.
However, choosing the effective strong coupling constants
as $\kappa_u = 0.01$ ($\Lambda=1$ TeV) , $f_u = 1/ \sqrt{2}$ and $h_u =  1/ \sqrt{2}$
and for a CM energy of 7 TeV,
the pure FCNC cross section is 8.718 pb, the interference 
term is 1.205 pb while the SM contribution amounts to
only 0.018 pb. Hence, the SM contributions can be safely 
neglected.

\begin{figure}[h!]
\centering
\hspace{-1.cm}\includegraphics[width=3.0in,angle=0]{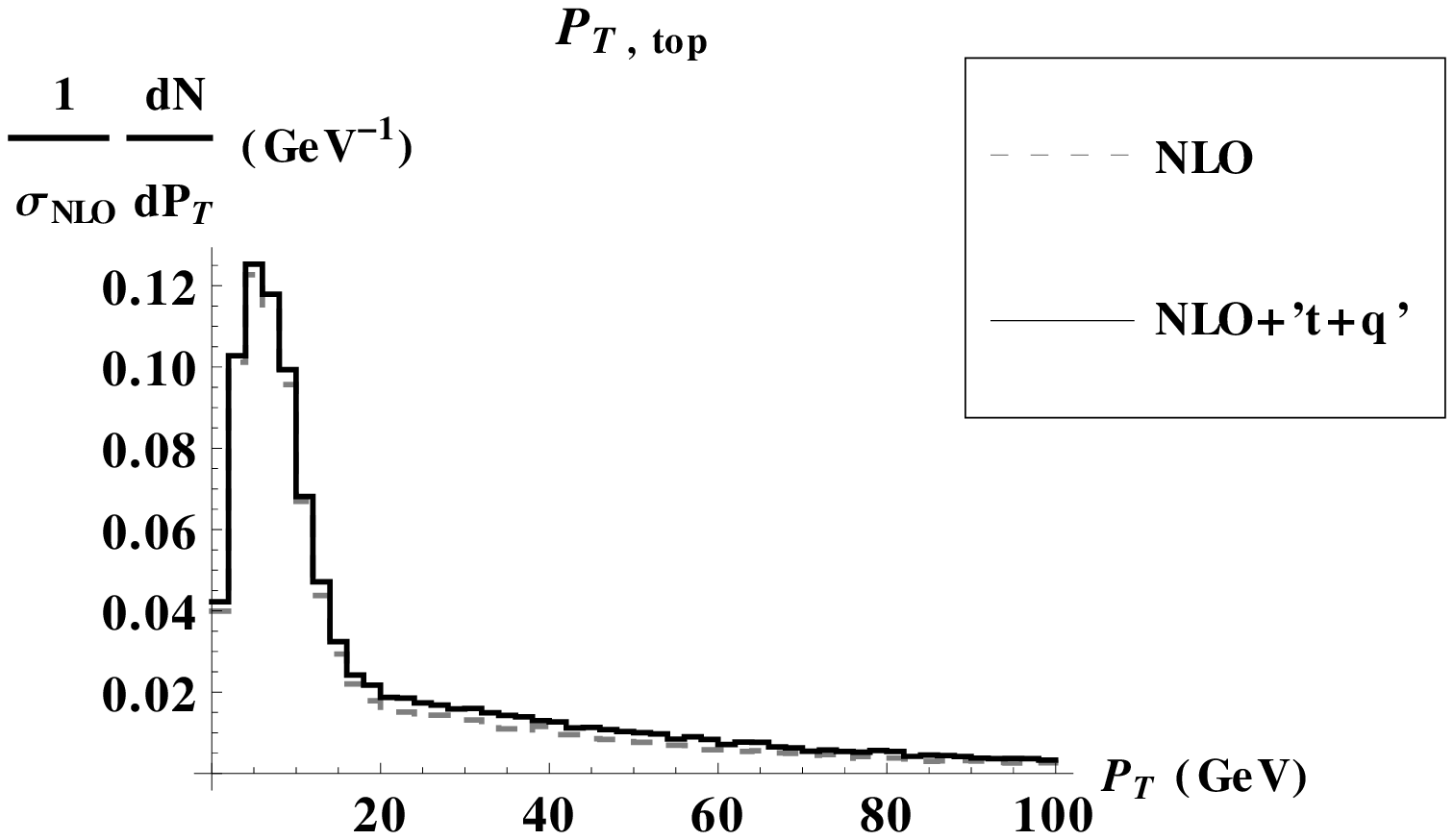}
\hspace{-.3cm}
\includegraphics[width=3.0in,angle=0]{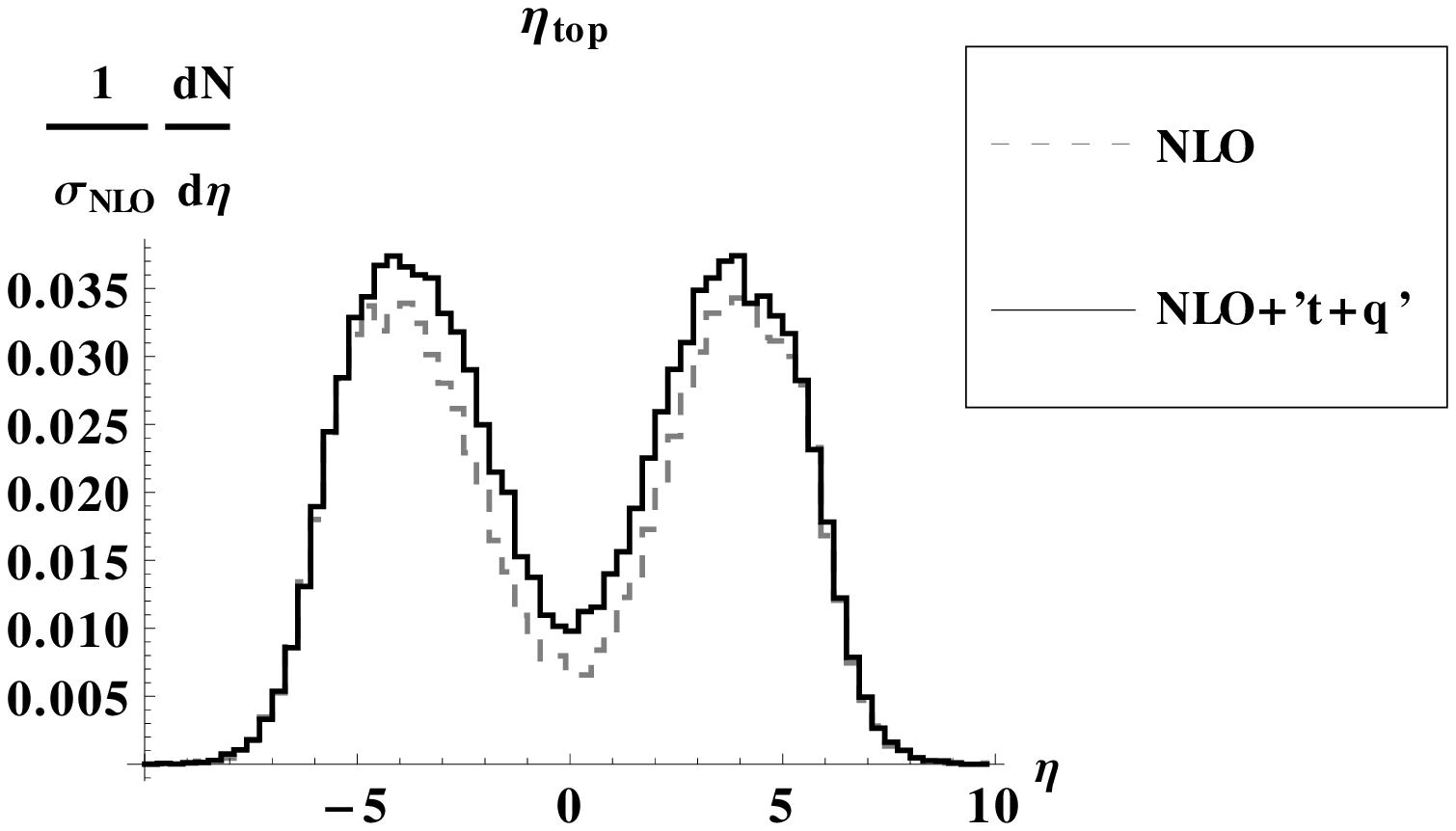}
\caption{$P_T$ (left)  and $\eta$ (right) distributions of the top quark for NLO direct top (solid line)
and NLO direct top plus $pp \to t q$ with $P_T^{match} = 10$ GeV and  jet $p_T > 10$ GeV. }
\label{fig:ptdtopnloq}
\end{figure}

In figure~\ref{fig:ptdtopnloq} we show the $P_T$ and $\eta$ distributions
for the direct top at NLO summed with $pp \to tq$ for a $P_T^{match} = 10$ GeV
and the jet $p_T > 10$ GeV. It is clear that the shape of the distributions do not change much with 
the inclusion of the $pp \to tq$ process but still the $pp \to tq$ process
gives a contribution of the order of 10 \% to the total cross section of the inclusive top
production at the LHC at $\sqrt{s} = 7$ TeV.

\section{Single top beyond the strong FCNC operators}

In the previous sections we have discussed NLO direct top
and $t+q$ production when only the strong FCNC operator
is considered. We note that the leading order contribution
to direct top does not receive contributions from other operators.
Therefore, the NLO calculation is again performed with only
the strong FCNC operator. As long as no excess is found 
at the LHC in the single top channel, the procedure described in the previous section
gives us the best possible bound on the anomalous strong FCNC 
coupling when all other operators are discarded.

In the hard $P_T$ region, the process $pp \to t + jet$ gets
contributions from the complete set of independent operators.
As these operators are independent from each other (and 
therefore so are the respective coupling constants) the interference
terms between strong and electroweak or 4F could 
be sizeable.
If an excess is found in the single top channel, one has to take 
into account all possible contributions 
from the remaining operators. A thorough analysis of 
the distributions of each individual operator will help us
understand which operators could be important for 
a given experimental analysis. Moreover, even if an excess is not seen 
in the single top channel, dedicated analysis could most
probably help constraining definite sets of operators.

\begin{figure}[h!]
\centering
\hspace{-1.cm}\includegraphics[width=3.0in,angle=0]{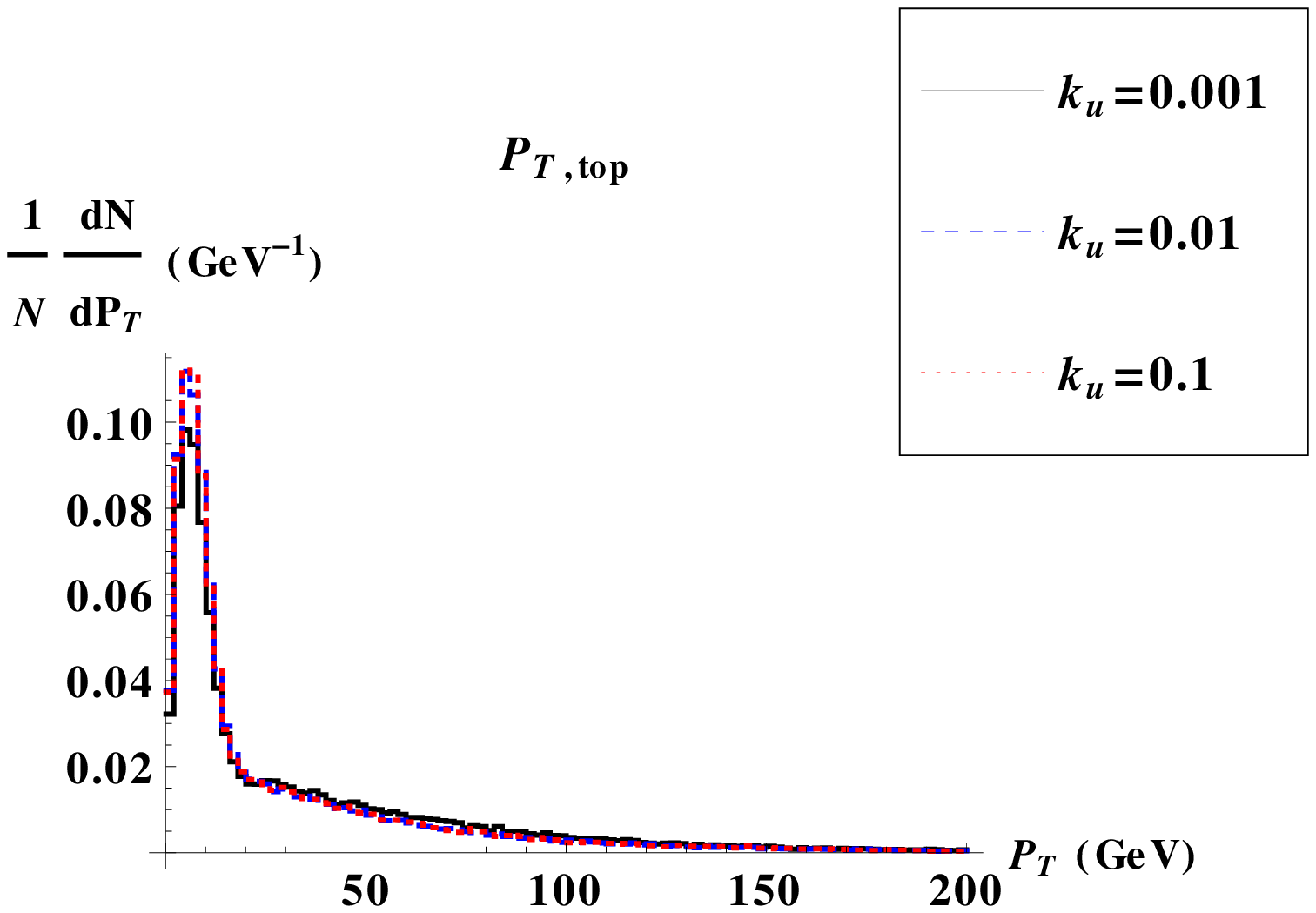}
\hspace{-.3cm}
\includegraphics[width=3.0in,angle=0]{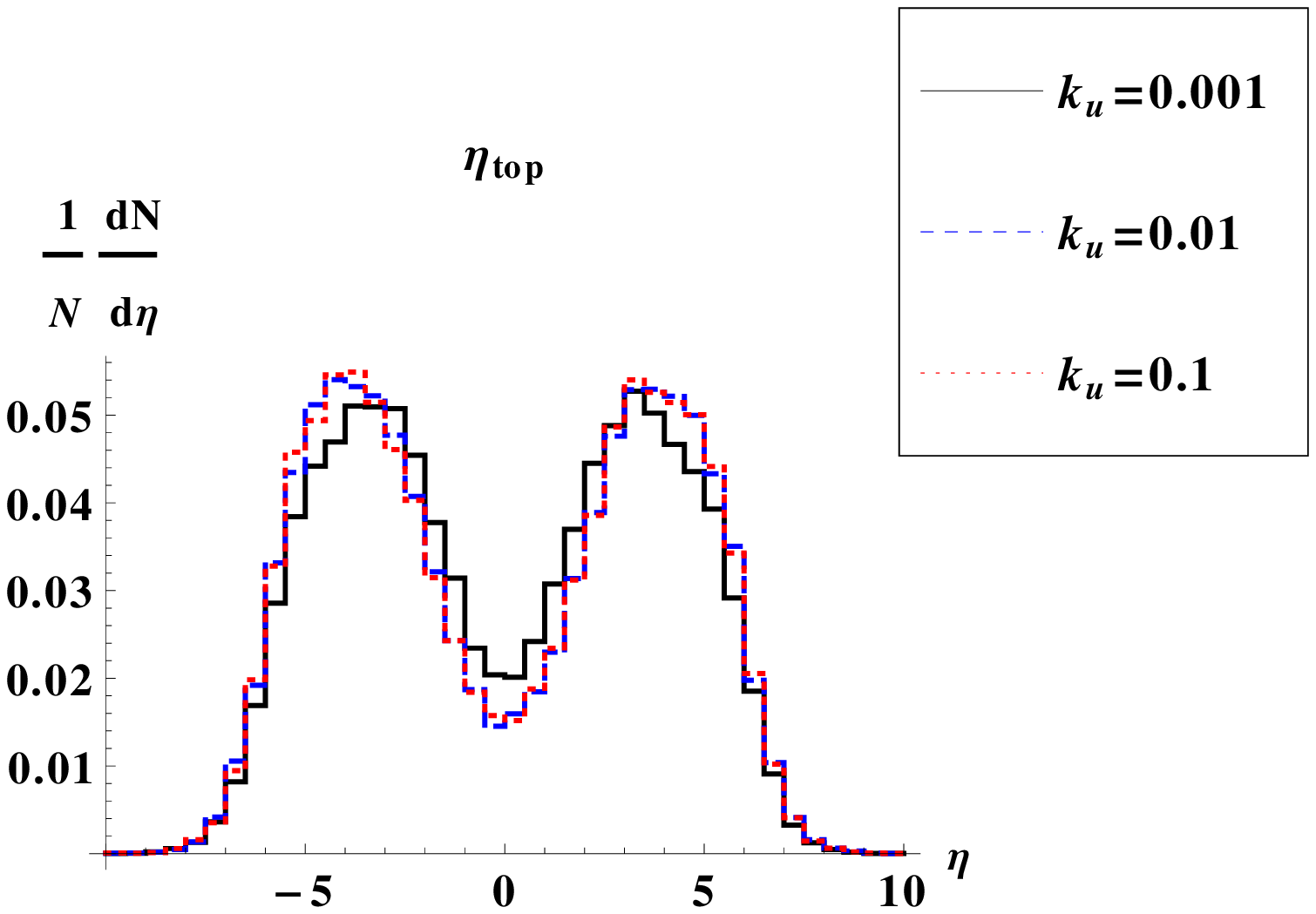}
\caption{$P_T$ (left)  and $\eta$ (right) distributions of the top quark when only the
strong operator is turned on with $P_T^{cut} = P_T^{match}= 10$ GeV. Process considered
is direct top at NLO plus $pp \to tq$ for $\sqrt{s} = 7$ TeV and three values of $\kappa_u$ with $\Lambda=1$ TeV. }
\label{fig:compare1}
\end{figure}

We start by considering the strong operator. When all other operators are turned off
the $P_T$ and $\eta$  distributions have a very mild dependence on the strong
coupling constant $\kappa$ ($\kappa_u$ to be more precise, and
we have set $\kappa_c =0$). This is shown in figure~\ref{fig:compare1}  where the 
$P_T$ (left)  and $\eta$ (right) distributions of the top quark are shown
for three values of $\kappa$, 0.001, 0.01 and 0.1 and $\Lambda=1$ TeV. The process is 
direct top NLO plus $pp \to tq$ for $\sqrt{s} = 7$ TeV and $P_T^{cut} = P_T^{match}= 10$ GeV.
It is clear that the shape of the distributions does not vary much making it possible
to perform the analysis for one constant and then to extract a bound on the 
strong operator.

\begin{figure}[h!]
\centering
\hspace{-1.cm}\includegraphics[width=3.0in,angle=0]{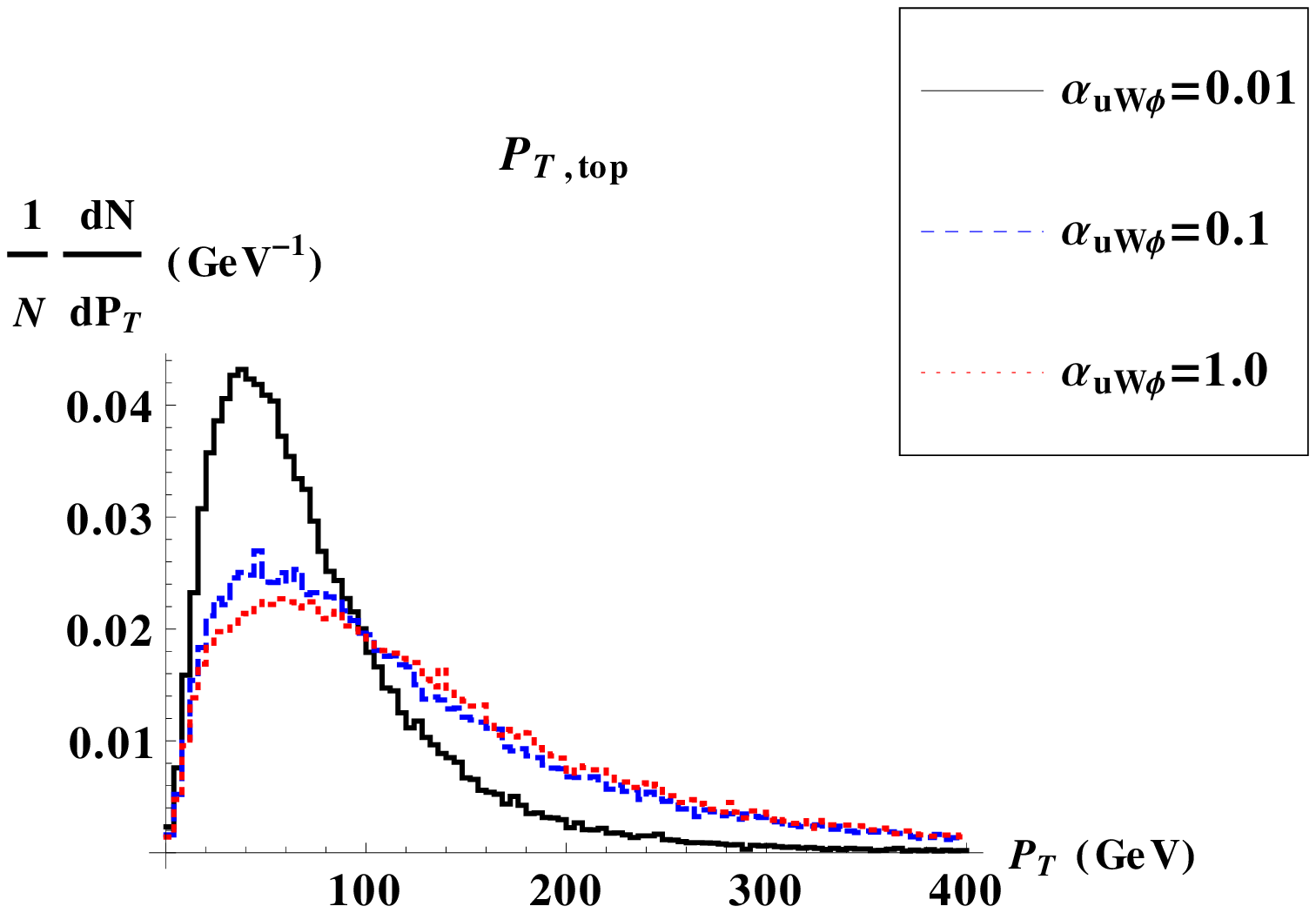}
\hspace{-.3cm}
\includegraphics[width=3.0in,angle=0]{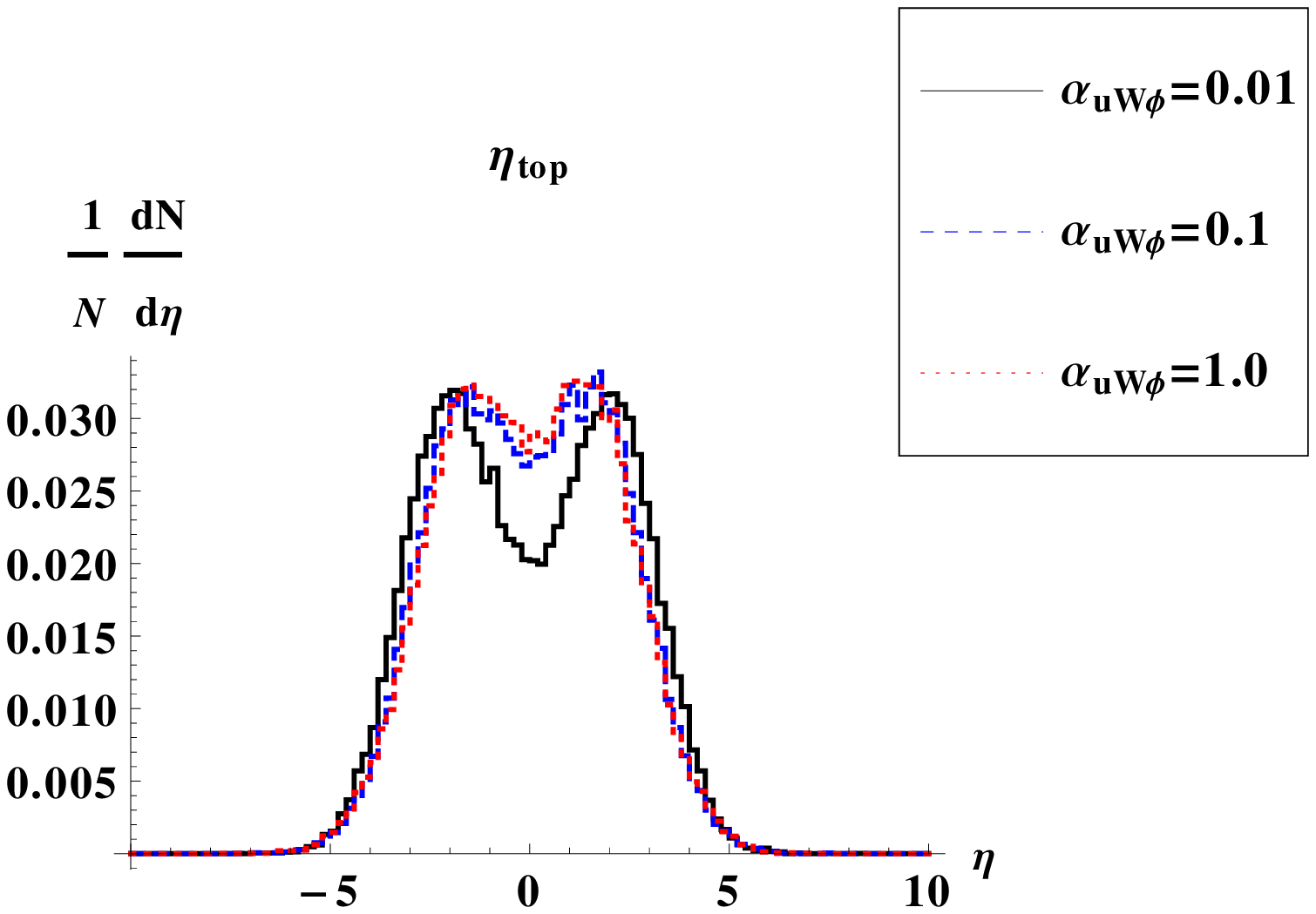}
\caption{$P_T$ (left)  and $\eta$ (right) distributions of the top quark when just
one electroweak operator, $O_{u W \phi}$, is turned on. 
The process is $pp \to tq$ for $\sqrt{s} = 7$ TeV and $P_T^{cut} = 10$ GeV.
}
\label{fig:compare2}
\end{figure}

We now move to the study of the electroweak operators.
We first consider only one operator $O_{u W \phi}$ turned on.
In figure~\ref{fig:compare2} we present the
$P_T$ (left)  and $\eta$ (right) distributions of the top quark for 
 three values of $\alpha_{u W \phi}$, 0.01 , 0.1 and 1 and $\Lambda=1$ TeV.
As $\alpha_{u W \phi} \to 0$
we recover the pure SM contribution of electroweak origin.
The SM cross section for this process and for $7$ TeV
is $\sigma = 0.019$ pb while the total cross section
for $\alpha_{u W \phi}= $ 0.01, 0.1 and 1 are $\sigma = 0.0020$ pb, 0.148 pb and 12.4 pb
respectively. Therefore the different shapes of the $P_T$ and $\eta$
distributions are due to the interference with the SM contribution.
When $\alpha_{u W \phi}= $  1, the total cross section is almost
100 times larger than its pure SM counterpart. Therefore, this
value shows how the distribution behaves when the SM contribution is negligible.

This kind of behaviour can occur for any operator on the list, provided
that the coupling constants are such that SM and FCNC cross sections are
of the same order of magnitude. Any deviation relative to the SM showing up in the distributions 
could mean an interference
with one or more operators. Understanding the different distribution requires 
dedicated studies with no assurance however that the responsible
operators could be identified. One should emphasise that a thorough
study of the $P_T$ and $\eta$ distributions of the top quark could help identifying classes
of FCNC operators.
A similar discussion applies
to the 4F operators case.

One should also note that $pp \to tq$ does not include direct top,
because the strong operator is turned off. Contrary to strong
operator scenario, in this case the distributions change with
the value of the electroweak constant. Therefore, any bound
based on the production process has to take into account
that different coupling constants can lead to different
distributions.

\begin{figure}[h!]
\centering
\hspace{-1.cm}\includegraphics[width=3.0in,angle=0]{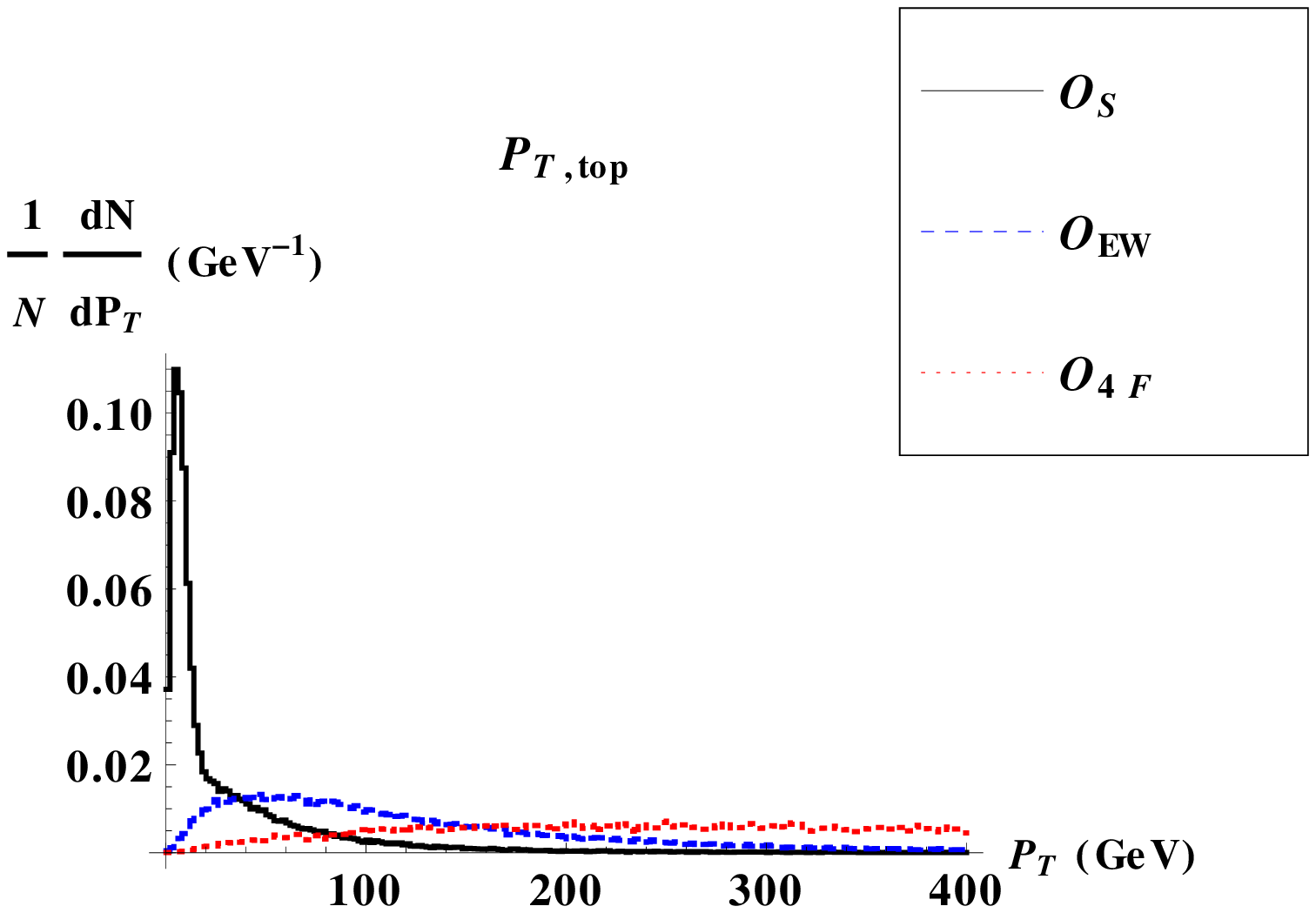}
\hspace{-.3cm}
\includegraphics[width=3.0in,angle=0]{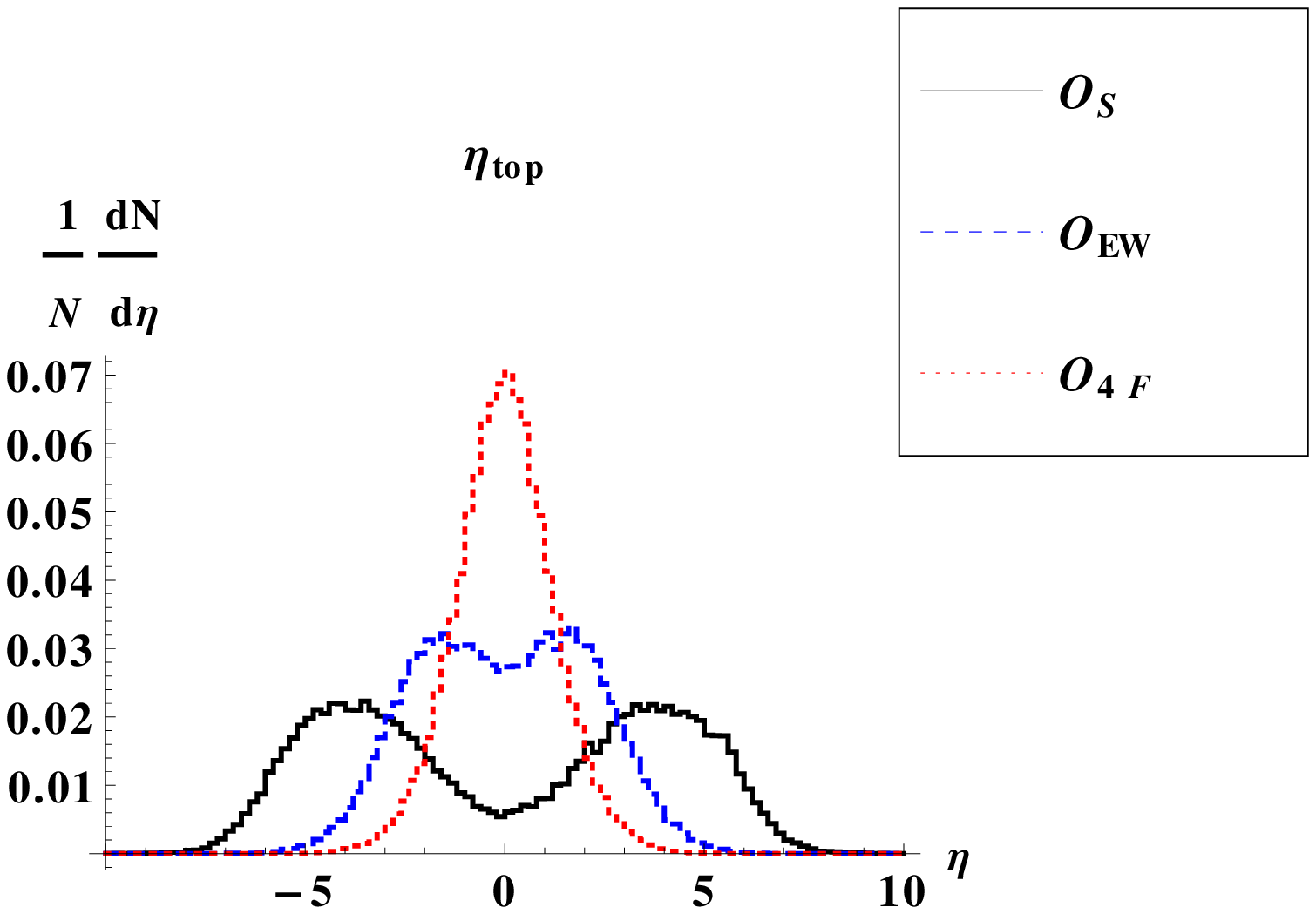}
\caption{$P_T$ (left)  and $\eta$ (right) distributions of the top quark when just one operator is taken non-zero at a time. We
compare the distributions of the strong FCNC operator with one electroweak, $O_{u W \phi}$, and one 4F operator. 
The process is $pp \to tq$ for $\sqrt{s} = 7$ TeV and $P_T^{cut} = 10$ GeV.}
\label{fig:compare3}
\end{figure}

Having studied the distributions of a definite operator representative of each class,
we will now perform a comparison between classes.
In figure~\ref{fig:compare3} we present the $P_T$ (left)  and $\eta$ (right) distributions of the top quark
when just one operator is taken non-zero at a time. We
compare the distributions of the strong FCNC operator with one electroweak operator (with coupling constant $\alpha_{u W \phi}$),
and one 4F operator, $(\bar{u}  \, \gamma_{\mu}   \, \gamma_{L} \, u) \,  (\bar{u}  \,\gamma^{\mu}   \, \gamma_{L}  \, t) $.
It is clear from the figure that
the distributions can be quite different and therefore distinguishable to some extent. The ability
to distinguish the different operators depends heavily on the relative values
of the coupling constants. If an excess in single top production is seen we can 
try to understand its origin by looking at all possible distributions. However,
this will always be a hard task because different operators give similar distributions
and therefore only very particular scenarios can be probed.

\section{Conclusions}

We have presented a new generator for the study of FCNC top interactions. 
The generator MEtop comes with different packages, 
each with a subset of a complete set of dimension
six operators. At the moment MEtop can generate events
for direct top and for top plus jet production, where the jet
can be any quark other then the top or it can be a gluon.

The direct top production process is implemented at NLO using
an effective NLO approximation. Also, the inclusive contribution
to direct top coming from $pp \to tq$ can be included
in the event generation.
We have shown that the top $P_T$ and $\eta$ distributions
show clear differences when the events are generated at LO
or at NLO. Therefore, the use of a constant K-factor does not provide an 
 accurate description of direct top production at NLO. We conclude that
 a new experimental analysis is in order to improve the constraints 
 on the strong FCNC coupling constants. The inclusion of
 the inclusive process $pp \to tq$ will further improve this bound.
We note that a detailed
study of the $P_T$ and $\eta$ distributions of the top quark could help identifying classes
of FCNC operators .

At LO, the contributions stemming from
the different operators can be compared in the single top production process.
In particular, 4F operators can be for the first time constrained
at hadron colliders. Constraining the 4F operators
can help us understand their role in the asymmetry 
measured at the Tevatron.

The bounds on $BR (t \to u (c) Z)$ and $BR (t \to u (c) \gamma)$ 
are obtained in the process $pp \to t \bar t$ where one of the
top-quarks decays as $t \to b W$ while the other decays as
$t \to u (c) Z$ or $t \to u (c) \gamma$. This means that
all electroweak FCNC couplings always appear in the same combination.
With MEtop we are able to look for distribution that isolates
each electroweak FCNC operator. This way more detailed
information can be obtained about each operator. 

New final states with FCNC contributions, like for instance $pp \to tW$~\cite{Etesami:2010ig} ,
are to be included in the next version of MEtop.

\vspace{0.5cm}
{\large{\textbf{Acknowledgments}}}

\vspace{0.5cm}
\noindent
We thank A. Belyaev,  A. Semenov,  T. Sj\"ostrand and Li Lin Yang for discussions.
We thank C. Friedrich and D. Hirschbuehl for their
help in the debugging and cross-check
with existing generators like PROTOS
and also for their support.

The work of R.C., A.O., R.S. and M.W.
is partially supported in part by the Portuguese
\textit{Funda\c{c}\~{a}o para a Ci\^{e}ncia e a Tecnologia} (FCT)
under contracts CERN/FP/123619/2011 and PTDC/FIS/117951/2010. 
R.S. is also partially supported by an
FP7 Reintegration Grant, number PERG08-GA-2010-277025
and by PEst-OE/FIS/UI0618/2011. M.W. is also supported by
FCT under contract SFRH/BD/45041/2008.

\appendix

\section{Effective couplings translation}

The representation of the effective coupling constants for each operator is arbitrary.  In this appendix
we will relate the two most common representations of the strong FCNC coupling constants 
appearing in the literature. The translation is simple and our goal is to clarify the 
relation between experimental bounds (and theoretical bounds as well) present in the literature.
When the effective strong operator comes from a dimension six effective lagrangian it is usually represented
in the form
\begin{equation}
{\cal L}_{ij}^{d6} \, = \, \frac{\alpha_{ij}}{\Lambda^2} \bar{q}^{i}_L \, \lambda^{a} \, \sigma^{\mu\nu} \, u^{j}_R \, \tilde{\phi} \, G^{a \mu \nu}  \,  + h.c.,
\label{eq:opap1}
\end{equation}
where $\Lambda$ is the scale of new physics. This operator can be written as a dimension five-like operator when the scalar field,
$\tilde{\phi}$ is replaced by $( v/\sqrt{2} \, \, \, \, \,   0)$, resulting in
\begin{equation}
{\cal L}_{ij}^{d6} \, = \,\frac{v}{\sqrt{2}} \frac{\alpha_{ij}}{\Lambda^2} \bar{u}^{i}_L \, \lambda^{a} \, \sigma^{\mu\nu} \, u^{j}_R \, G^{a \mu \nu}  \,  + h.c. \, \, .
\label{eq:opap2}
\end{equation}
There are in principle four complex constant involved, $\alpha_{it} $ and $\alpha_{ti}$ with $i=u,c$ in a total of four degrees of freedom.
On the other hand, several authors adopt to write the same strong operator as a dimension five effective operator. In this
case it is usually written as 
\begin{equation}
{\cal L}_{ij}^{d5} \, = \, - g_S \frac{\kappa_{ij}}{\Lambda} \bar{u}^{i} \, \lambda^{a} \, \sigma^{\mu\nu} \, (f_{ij} + i h_{ij} \gamma_5)\, u^{j} \, G^{a \mu \nu}  \,  + h.c.,
\label{eq:opap3}
\end{equation}
 where  $g_S$ is the strong coupling constant, $\kappa_{ij}$ is taken as real and positive, $f_{ij}$ and $h_{ij}$ are complex and 
$|f_{ij}|^2 + |h_{ij}|^2 = 1$. In both cases $Tr[\lambda_a \lambda_b] = \delta_{ab}/2$, and the vacuum expectation value is  $v=246$ GeV. Note
that in reference~\cite{buch} (as in other references) the vacuum expectation value is defined as $v = 246/\sqrt{2}$ GeV. 
The constant $\kappa_{ij}$ is real and only two constants are needed, one for
each light flavour; the same is true for the complex constants $f_{ij}$ and $h_{ij}$. Therefore we can use just the light quark index
to represent those coupling constants
\begin{equation}
{\cal L}_{i}^{d5} \, = \, - g_S \frac{\kappa_{i}}{\Lambda} \bar{u}^{i} \, \lambda^{a} \, \sigma^{\mu\nu} \, (f_{i} + i h_{i} \gamma_5)\, u^{j} \, G^{a \mu \nu}  \, + h.c. ,
\label{eq:opap4}
\end{equation}
and in this case the index $j$ refers to the top quark.
It is now straightforward to find the relation between the two set of operators
\begin{eqnarray}
&& \alpha_{it} \, = \, - \sqrt{2} \, g_S \, \frac{\Lambda}{v}\, \kappa_{i} \, (f_{i} + i h_{i})              \nonumber \\ 
&& \alpha_{ti} \, = \, - \sqrt{2} \, g_S \, \frac{\Lambda}{v}\, \kappa_{i} \, (f_{i}^* + i h_{i}^*)
\end{eqnarray}
with $i=u,c$. In most cases all constants are taken as real. This means that equation~\ref{eq:opap4} can be written as
\begin{equation}
{\cal L}_{i}^{d5} \, = \, - g_S \frac{\kappa_{i}}{\Lambda} \bar{u}^{i} \, \lambda^{a} \, \sigma^{\mu\nu} \, (f_{i} + h_{i} \gamma_5)\, u^{j} \, G^{a \mu \nu}  \,  + h.c.,
\label{eq:opap5}
\end{equation}
and consequently
\begin{eqnarray}
&& \alpha_{it} \, = \, - \sqrt{2} \, g_S \, \frac{\Lambda}{v}\, \kappa_{i} \, (f_{i} + h_{i})              \nonumber \\ 
&& \alpha_{ti} \, = \, - \sqrt{2} \, g_S \, \frac{\Lambda}{v}\, \kappa_{i} \, (f_{i} - h_{i})
\end{eqnarray}
with $i=u,c$, all constants are now real and $|f_{i}|^2 + |h_{i}|^2 = 1$.


\section{The complete dimension six lagrangian for single top production}

As previously discussed, the number of effective dimension six operators is huge. 
Therefore, no meaningful analysis is possible when all operators are considered
simultaneously. Any subset of operators, however small, does not simplify much
the task of obtaining information about each individual operator.
In order to perform any relevant study involving all different types
of 4F operators, we have built a 4F subset to be used as a basis 
for our study. Since we are dealing with hadron colliders, our first simplification
is to consider only processes initiated by u-quarks. This is equivalent
to say that the coupling constants in the 4F sector are all of the same order. 
Using references~\cite{buch, Grzadkowski:2010es, AguilarSaavedra:2010zi} 
we extract nine 4F effective operators that could contribute to single top production
at hadron colliders. This set can be written as
\begin{eqnarray}
{\cal O}_{q q}^{ijkl}&=&\frac{1}{2} ({\bar q}_{L i}\, \gamma^\mu\, q_{L j})\; ({\bar q}_{Lk}\, \gamma_\mu\, q_{Ll}),\\
{\cal O}_{q q'}^{ijkl}&=&\frac{1}{2} ({\bar q}_{L ia}\, \gamma^\mu\, q_{L j b})\; ({\bar q}_{Lkb}\, \gamma_\mu\, q_{Lla}),\\
{\cal O}_{u u}^{ijkl}&=&\frac{1}{2} ({\bar u}_{R i}\, \gamma^\mu\, u_{R j})\; ({\bar u}_{Rk}\, \gamma_\mu\, u_{Rl}),\\
{\cal O}_{u d}^{ijkl}&=&({\bar u}_{R i}\, \gamma^\mu\, u_{R j})\; ({\bar d}_{Rk}\, \gamma_\mu\, d_{Rl}) ,\\
{\cal O}_{u d'}^{ijkl}&=&({\bar u}_{R i a}\, \gamma^\mu\, u_{R j b})\; ({\bar d}_{Rkb} \gamma_\mu\, d_{Rla}),\\
{\cal O}_{q u}^{ijkl}&=& ({\bar q}_{Li}\, u_{Rj})\; ({\bar u}_{Rk}\, q_{Ll}),\\
{\cal O}_{q u'}^{ijkl}&=&({\bar q}_{Lia}\, u_{Rjb})\; ({\bar u}_{Rkb} q_{Lla}),\\
{\cal O}_{q d}^{ijkl}&=&({\bar q}_{L i}\, d_{Rj})\; ({\bar d}_{Rk}\, q_{L l}),\\
{\cal O}_{q d'}^{ijkl}&=&({\bar q}_{L i a}\, d_{Rjb})\; ({\bar d}_{Rkb} q_{L l a}), \\
\end{eqnarray}
where $i,j,k,l=1,2,3$ are flavour indices and the sub-indices $a$ and $b$ indicate the contraction
of color indices whenever this pairing is different from the one in spinor contraction. 
In \cite{AguilarSaavedra:2010zi} this set of operators was simplified to the one presented 
in table~\ref{table:opqqtq}. 
\begin{table}[h!]
\begin{center}
    \begin{tabular}{  c | c  }
    \hline \hline
	$\frac{1}{2}(\alpha_{qq}^{kji3}+\alpha_{qq'}^{ijk3})(\bar{u}_{Lk} \gamma^\mu u_{Lj})(\bar{u}_{Li} \gamma_\mu t_{L})$ & $-\frac{1}{2}\alpha_{qu'}^{k3ij}(\bar{u}_{Lk} \gamma^\mu u_{Lj})(\bar{u}_{Ri} \gamma_\mu t_{R})$ \\
    \hline
	$-\frac{1}{2}\alpha_{qu'}^{ijk3}(\bar{u}_{Rk} \gamma^\mu u_{Rj})(\bar{u}_{Li} \gamma_\mu t_{L})$ & $\frac{1}{2}\alpha_{uu}^{kji3}(\bar{u}_{Rk} \gamma^\mu u_{Rj})(\bar{u}_{Ri} \gamma_\mu t_{R})$ \\
    \hline
	$-\frac{1}{2}\alpha_{qu}^{k3ij}(\bar{u}_{Lka} \gamma^\mu u_{Ljb})(\bar{u}_{Rib} \gamma_\mu t_{Ra})$ & $-\frac{1}{2}\alpha_{qu}^{ijk3}(\bar{u}_{Rka} \gamma^\mu u_{Rjb})(\bar{u}_{Lia} \gamma_\mu t_{La})$ \\
    \hline \hline
    \end{tabular}
\end{center}
\caption{4F operators contributing to single top production with parton level processes of the type $u (\bar u) u (\bar u) \to t (\bar t) u (c \, \bar c \, \bar u) $.}
\label{table:opqqtq}
\end{table}
There is a total of 24 different combinations coming from setting 
two of the indices $i, j, k$ equal to 1 while the remaining one 
is set to 1 or 2. By forcing the initial state to be composed
of  u-quarks only, we can further reduce the number of
operators to 12 (we only allow for one "FCNC-current",  ($\bar{u} \Gamma  t$)
or ($\bar{c} \Gamma  t$), where $\Gamma$ stands for a generic Lorentz structure).
Therefore the final 12 independent 4F operators are the ones obtained
by setting $k=j=1$ and $i=1,2$. Because the operators had to be rearranged
to allow for the implementation in LanHEP (see discussion below), the final lagrangian is then written as 
\begin{eqnarray}
{\cal L}_{qq,qg,gg \rightarrow t\, {\bar q}}&=& \frac{1}{\Lambda^2} \, \sum_{\substack{i,j=1,3 \\ or \\ i,j=2,3 \\ i\neq j }} \Big( 
\alpha_{uG \phi }^{ij} {\cal O}_{u G \phi}^{ij}
+\alpha_{uW \phi }^{ij} {\cal O}_{uW \phi}^{ij}
+\alpha_{u B \phi}^{ij}\, {\cal O}_{u B \phi}^{ij} 
+\alpha_{\phi u}^{ij}\, {\cal O}_{\phi u}^{ij}+\alpha_{\phi q}^{(3,ij)}\, {\cal O}_{\phi q}^{(3,ij)}  
\nonumber\\
&& 
+\alpha_{\phi q}^{(1,ij)} \,  {\cal O}_{\phi q}^{(1,ij)} 
+\alpha_{u \phi} \, {\cal O}_{u\phi}^{ij}  \Big) + \frac{1}{\Lambda^2} \, {\cal L}_{4fu}+
 \frac{1}{\Lambda^2} \, {\cal L}_{4fc}
\end{eqnarray}
where ${\cal L}_{4fu}$ is the 4F lagrangian for anomalous top-up coupling
\begin{eqnarray}
{\cal L}_{4fu}&=&\frac{1}{2}(\alpha_{qq}^{1113}+\alpha_{qq'}^{1113}) ({\bar u}_L\, \gamma^\mu u_L) ({\bar u}_{L} \gamma_\mu t_L) 
\nonumber\\
&&-\frac{1}{2} (\alpha_{qu'}^{1311}+\frac{1}{3} \alpha_{qu}^{1311}) ({\bar u}_L\gamma^\mu u_L) ({\bar u}_{R} \gamma_\mu t_R)
\nonumber\\
&&-\frac{1}{2} (\alpha_{qu'}^{1113}+\frac{1}{3} \alpha_{qu}^{1113}) ({\bar u}_R\gamma^\mu u_R) ({\bar u}_{L} \gamma_\mu t_L)
\nonumber\\
&&+\frac{1}{2} \alpha_{uu}^{1113} ({\bar u}_R\gamma^\mu u_R) ({\bar u}_{R} \gamma_\mu t_R)
\nonumber\\
&&-\frac{1}{4} \alpha_{qu}^{1311} ({\bar u}_{L}\gamma^\mu \lambda^a u_{L}) ({\bar u}_{R} \gamma_\mu \lambda^a t_{R})
\nonumber\\
&&-\frac{1}{4} \alpha_{qu}^{1113} ({\bar u}_{R}\gamma^\mu \lambda^a u_{R}) ({\bar u}_{L} \gamma_\mu \lambda^a t_{L})
\label{eq:4Fu}
\end{eqnarray}
and  ${\cal L}_{4fc}$ is the 4F lagrangian for anomalous top-charm coupling
\begin{eqnarray}
{\cal L}_{4fc}&=&\frac{1}{2}(\alpha_{qq}^{1123}+\alpha_{qq'}^{2113}+\frac{1}{3}\alpha_{qq}^{2113}+\frac{1}{3}
\alpha_{qq'}^{1123}) ({\bar u}_L\, \gamma^\mu u_L) ({\bar c}_{L} \gamma_\mu t_L) 
\nonumber\\
&&-\frac{1}{2} (\alpha_{qu'}^{1321}+\frac{1}{3} \alpha_{qu}^{1321}) ({\bar u}_L\gamma^\mu u_L) ({\bar c}_{R} \gamma_\mu t_R)
\nonumber\\
&&-\frac{1}{2} (\alpha_{qu'}^{2113}+\frac{1}{3} \alpha_{qu}^{2113}) ({\bar u}_R\gamma^\mu u_R) ({\bar c}_{L} \gamma_\mu t_L)
\nonumber\\
&&+\frac{1}{2} (\alpha_{uu}^{1123}+\frac{1}{3}\alpha_{uu}^{2113}) ({\bar u}_R\gamma^\mu u_R) ({\bar c}_{R} \gamma_\mu t_R)
\nonumber\\
&&-\frac{1}{4} \alpha_{qu}^{1321} ({\bar u}_{L}\gamma^\mu \lambda^a u_{L}) ({\bar c}_{R} \gamma_\mu \lambda^a t_{R})
\nonumber\\
&&-\frac{1}{4} \alpha_{qu}^{2113} ({\bar u}_{R}\gamma^\mu \lambda^a u_{R}) ({\bar c}_{L} \gamma_\mu \lambda^a t_{L})
 \nonumber\\
&&+
\frac{1}{4}(\alpha_{qq}^{2113}+\alpha_{qq'}^{1123})({\bar c}_L \gamma^\mu \lambda^a t_L) ({\bar u}_L \gamma^\mu \lambda^a u_L)
\nonumber\\
&&+\frac{1}{4}\alpha_{uu}^{2113} ({\bar c}_R \gamma^\mu \lambda^a t_R) ({\bar u}_R \gamma^\mu \lambda^a u_R) \nonumber\\
&&+(\frac{1}{3}\alpha_{qu'}^{2311}+\alpha_{qu}^{2311}) ({\bar c}_L t_R) ({\bar u}_R u_L)
+\frac{1}{2}\alpha_{qu'}^{2311} ({\bar c}_L \lambda^a t_R) ({\bar u}_R \lambda^a u_L)\nonumber\\
&&+(\frac{1}{3}\alpha_{qu'}^{1123}+\alpha_{qu}^{1123}) ({\bar c}_R t_L) ({\bar u}_L u_R))+\frac{1}{2}\alpha_{qu'}^{1123} ({\bar c}_R \lambda^a t_L) ({\bar u}_L \lambda^a u_R).
\label{eq:4Fc}
\end{eqnarray}
The operators with the Gell-Mann matrices originate from re-writing the ones  where the quark colours indices were
explicitly summed. The inclusion of 4F operators in MEtop was done by  implementing 
the 4F effective lagrangian in LanHEP. All 4F operators in table~\ref{table:opqqtq} have four coloured particles
converging in one point which is a type of interaction LanHEP is not
able to handle automatically due to the complex color flow. 
Therefore, we had to implement these operators  using an auxiliary
field mechanism \cite{Belyaev:2005ew}, where the 4-color vertex is
replaced by 3-color vertices that when combined in s,t and u channels,
will reconstruct the 4-fermion interaction. These 3-color vertices are 
implemented by introducing the interaction terms in the initial lagrangian together
with a unit mass field with a point-like propagator. An example of how a lagrangian
is written is shown in equation
\ref{eqn:aux} with a vectorial auxiliary field
\begin{equation}
{\cal L}_{4F} = (\bar{\psi}_L^i \gamma^\mu \psi_L^j)(\bar{\psi}_L^k
\gamma_\mu \psi_L^l) \to (\bar{\psi}_L^i \gamma^\mu \psi_L^j) X_\mu +
X^\nu (\bar{\psi}_L^k \gamma_\nu \psi_L^l) + \frac{1}{2} X_\mu X^\nu
\label{eqn:aux}
\end{equation}
where $ \psi_L$ is a left-handed spinor and $X_\mu$ is a spin 1 field that does not propagate.

\section{Using MEtop}

\subsection{Installation}

MEtop is written in C and python and it generates events following the \textit{Les Houches Accord} format.
It can therefore be easily interfaced with PYTHIA or Herwig. In order to compile it,  
you need a C compiler~ \footnote{There is one file written in Fortran and therefore you also need a Fortran compiler.}
and python version 2.6 or later. To run the package you must additionally install
\begin{itemize}
\item Cuba Library version 3.0
\item LHAPDF version 5.8.6
\item Numpy version 1.3.0
\end{itemize}
The Cuba and LHAPDF library must be available through the library environment variable (for example).

To install MEtop you just have to execute "make" in the main directory.

\subsection{The generator}

\subsubsection{param.dat}

In MEtop all parameters are set in one file: "param.dat". Table \ref{tbl:param} summarizes the definition of each parameter.

\begin{center}
\begin{table}[h!]
\begin{center}
\begin{tabular}{ l | c  }
  \hline  \hline                     
   Mx & Particle's masses (x=u,d,c,s,b,top,e,$\mu$,$\tau$,W,Z,H)  \\
  wx & Particle's Widths (x=W,top,Z,H)  \\
  sx &   Values for CKM matrix elements (x=12,23,13)  \\
  SW &  $\sin \theta_W$ ( $\theta_W$ is the Weinberg angle)\\
  EE  &  Electromagnetic coupling constant \\
  cox & couplings of the x operator (x=1,2,...,9)\\
  fx,hx & Chirality parameters from operators co1 and co2\\
  Q   & Factorization scale \\
  miuR & Renormalization scale for Direct top at NLO \\
  L  & Energy scale \\
  ECM & Centre of mass Energy\\
  PTmatch & PT for matching \\
  PTmin & Cut in PT for LO $2 \rightarrow 2$ processes \\
  NEvnts & Number of events to generate \\
  pdf & PDF name according to LHAPDF \\
  pp  &  Type of collider: 1 for $pp$ and -1 for $p\bar{p}$ \\
  DecMod  & Turn on/off W decay modes\\
  SpCorr  & Turn on/off Spin Correlations\\
  ttbar  & $t$, $\bar{t}$ channel. 0-$t$ only;1-$\bar{t}$ only;2-$t$ and $\bar{t}$\\
  seed  & Turn random number seed\\
  \hline \hline  
\end{tabular}
\caption{Summary description of "param.dat" file.}
\label{tbl:param}
\end{center}
\end{table}
\end{center}

\subsubsection{Physical processes}

 In addition to the parameters defined in table~\ref{tbl:param} there are two more flags in  "param.dat" file: "cs" and "Process". 
 The first one dictates whether or not to calculate the cross sections and/or to generate events. 
 The second sets which physical process should be taken into account. If "cs" is set to 0, 
 the cross sections for all sub-processes defined by the  "Process" flag will be 
 calculated and no generation will be performed. The result will be stored in the 
 CS folder, in a csX.txt file, where X can be "Dtop","Gtop" and "Lqtop". If "cs" is set to 1, 
 only the event generation will be performed. In this case events are produced  according 
 to the calculated cross sections. After generation, the \textit{LHE} files will be stored in the Events 
 folder together with a file "runinfo.txt" which stores all information related to the event generation. 

top quark FCNC interactions were introduced in MEtop through an effective lagrangian.  Depending on which
operators are "turned on", different physics will be generated. Two different topologies are available: 
$2 \rightarrow 1 \rightarrow 3$ and $2 \rightarrow 2 \rightarrow 4$.~\footnote{
When "SpCorr" is set to 0, the top quark decay will \underline{not} be performed in MEtop, that is, 
the generated events will have the topology $2 \rightarrow 1$ and $2 \rightarrow 2$. 
In this case the spin correlations are lost.
}
The first one concerns 
"Direct top" production, and the second is related to "top+gluon" and "top+ light quark". 

\begin{center}
\begin{table}[h!]
\begin{center}
\begin{tabular}{ c | c | c }
  \hline  \hline 
Process Number &    Description          &   Comments  \\   \hline
1                        &    Direct top (LO)     &   Strong Op. only   \\
2                        &    top+gluon (LO)    &  Strong Op. only; set PTmin \\
3                        &    top+quark (LO)    &  All Op.; set PTmin \\
21                      &    Direct top (NLO)    &  Strong Op. only \\
22                      &    Direct top (NLO) + top+quark (LO)     &  All Op. \\
  \hline \hline  
\end{tabular}
\caption{Processes available in MEtop}
\label{tbl:proc}
\end{center}
\end{table}
\end{center}

Strong FCNC top interactions are included in MEtop through two equivalent
effective operators, one for the top and u-quark interaction, and the other
for the interaction of the top with a c-quark. 
In process 1, only the strong coupling constants are needed. Process 2 has the same effective
operators but due to the infrared divergences appearing in top+gluon production a cut 
in the top quark transverse momentum has to be set via the variable PTmin.
In process 3, top + light quark~\footnote{Here light quark stands for the set $u,\bar{u},d,\bar{d},c,\bar{c},s,\bar{s},b,\bar{b}$.}
production, all operators can contribute, strong, electroweak and 4F. It is now possible to choose which
operators to include. Again a value for PTmin has to be chosen.

Process 21 is inclusive direct top production at NLO and again only strong
operators intervene. The NLO result is obtained by a matching procedure 
(as described previously) which depends on one variable, PTmatch, to be
chosen by the user. The cross section results are written in three files:
 "csDtopLO.txt", the LO result for direct top, "csDtopNLO.txt", the NLO
increment relative to the LO result ($\sigma^{Total}_{NLO}-\sigma_{LO}^{Total}$),
and "csGtop.txt", the LO cross section for "top+gluon" process with a top
quark transverse momentum above "PTmatch". Therefore the variable "PTmin" is irrelevant
for this process.
After the generation, the results are stored in one file in the Events folder named 
"DtopNLO.lhe", 
containing  $2 \rightarrow 1 \rightarrow 3$ and the $2 \rightarrow 2 \rightarrow 4$ configurations. 
These events constitute the inclusive direct top NLO event generation, and must 
subsequently be showered by PYTHIA using the PT-ordered scheme, in order to complete the matching procedure.
Finally, with process 22, MEtop sums process 21 with process 3. The "PTmacth" variable 
plays the same role as in process 21 and "PTmin" will be the top transverse momentum cut, 
for the "top + quark" sub-processes.

\subsection{Running MEtop}

To run the package you just have to execute the command "./run.py" in the main directory. 
Care should be taken when changing the values of the physical parameters and/or the process 
you wish to calculate. In such cases you must always recalculate the value of the cross section. In addition, 
if you change the process used for the generation, you must be sure that all cross sections pertaining the 
new process are calculated beforehand. This is mandatory because the generation is done using 
the cs*.txt files saved in the CS folder.

\subsection{Available Model files}

At the moment there are three different packages available in MEtop, with a different set of effective operators. 
The main reason to have the different packages is to make the generation of events faster. The lightest
version includes only the strong sector. Then there are two other versions one with strong plus 
electroweak operators and the other one with strong plus 4F operators.

In equation \ref{eqn:package1} we present the strong FCNC lagrangian as it is written in the  
package "MEtop\_S\_vxx.tar.gz"
\begin{eqnarray}
{\cal L}_{S}&=& co1 \,  {\cal O}_{u G } + co2\,  {\cal O}_{c G } + h.c.
\label{eqn:package1}
\end{eqnarray}
with
\begin{eqnarray}
 {\cal O}_{u G } = i \frac{g_s}{\Lambda} \bar{u} \lambda^a \sigma^{\mu\nu} (fu+hu \, \, \gamma_5) t  G_{\mu\nu}^a \quad, 
 \quad {\cal O}_{c G } = i \frac{g_s}{\Lambda} \bar{c} \lambda^a \sigma^{\mu\nu} (fc+hc \, \, \gamma_5) t  G_{\mu\nu}^a
\end{eqnarray}
and $co1$, $co2$, $fu$, $hu$, $fc$, $hc$ are real constants to be chosen in the file param.dat. The constants
$f_i$, $h_i$ allow the choice of different chiralities while $co_i$ are overall normalization constants. Although
it usually considered that $f_i^2 + h_i^2 =1$, this relation has to be implemented by the user by a judicious 
choice of parameters $f_i$ and $h_i$.

The package "MEtop\_SEW\_vxx.tar.gz" contains the strong and electroweak sectors.
The lagrangian introduced in this package is
\begin{eqnarray}
{\cal L}_{SEW}&=& {\cal L}_{S}+ \frac{1}{\Lambda^2} \, \sum_{\substack{i,j=1,3 \\  i\neq j }} \Big( 
\alpha_{uW}^{ij} {\cal O}_{uW}^{ij}
+\alpha_{u B \phi}^{ij}\, {\cal O}_{u B \phi}^{ij} 
+\alpha_{\phi u}^{ij}\, {\cal O}_{\phi u}^{ij}+\alpha_{\phi q}^{(3,ij)}\, {\cal O}_{\phi q}^{(3,ij)}  
\nonumber\\
&& 
+\alpha_{\phi q}^{(1,ij)} \,  {\cal O}_{\phi q}^{(1,ij)} 
+\alpha_{u \phi} \, {\cal O}_{u\phi}^{ij}  \Big) 
\label{eqn:package2}
\end{eqnarray}
where the electroweak operators are
\begin{eqnarray}
\nonumber
{\cal O}_{u\phi}^{ij} &=& (\phi^\dagger \phi)\, ({\bar q}_{L i}\, u_{R j}\, {\tilde \phi})
\quad \quad \quad \quad \quad, \quad \quad
{\cal O}_{\phi q}^{(1,ij)} =  i\, (\phi^\dagger\, D_\mu\, \phi)\, ({\bar q}_{L i}\, \gamma^\mu\, q_{L j})\\ 
\nonumber
{\cal O}_{\phi q}^{(3,ij)}&=& i\, (\phi^\dagger\, D_\mu\, \tau^I\, \phi)\, ({\bar q}_{L i}\, \gamma^\mu\,\tau^I\, q_{L j})
\quad , \quad
{\cal O}_{\phi u}^{ij} = i\, (\phi^\dagger\, D_\mu\, \phi)\, ({\bar u}_{R i}\, \gamma^\mu\, u_{R j})\\
\nonumber
{\cal O}_{u W}^{ij}&=&({\bar q}_{L i}\, \sigma_{\mu\nu}\, \tau_I\, u_{R j})\, \tilde{\phi}\, W_{\mu\nu}^I 
 \quad \quad \quad , \quad \quad
{\cal O}_{u B \phi}^{ij}=({\bar q}_{L i}\, \sigma_{\mu\nu}\, u_{R j})\, \tilde{\phi}\, B_{\mu\nu}
\end{eqnarray}
and all coupling constants are real. In param.dat all coupling constant have the form $co_i$.
The relation between the coupling constants presented in equation \ref{eqn:package2} and the $co_i$
parameters to be chosen in param.dat is presented in table~\ref{tbl:dicLSEW}. 
\begin{table}[h!]
\begin{center}
    \begin{tabular}{  c | c | c | c }
    \hline \hline
	$co3 \rightarrow \alpha_{uW}^{ut}$ & $co4 \rightarrow \alpha_{uW}^{tu}$  & $co5 \rightarrow \alpha_{u B \phi}^{ut}$  & $co6 \rightarrow \alpha_{u B \phi}^{tu}$ \\
    \hline
	$co7 \rightarrow \alpha_{\phi u}^{ut}$  &   $co8 \rightarrow \alpha_{\phi u}^{tu}$  & $co9 \rightarrow \alpha_{\phi q}^{(3,ut)} $ &  $co10 \rightarrow \alpha_{\phi q}^{(3,tu)}$ \\
    \hline
	 $co11 \rightarrow \alpha_{\phi q}^{(1,ut)}$ & $co12 \rightarrow \alpha_{\phi q}^{(1,tu)}$ & $co13 \rightarrow \alpha_{u \phi}^{ut}$ &  $co14 \rightarrow \alpha_{u \phi}^{tu}$  \\
    \hline \hline
    \end{tabular}
\end{center}
\caption{Coefficient dictionary for ${\cal L}_{SEW}$ }
\label{tbl:dicLSEW}
\end{table}

Finally, the file "MEtop\_S4F\_vxx.tar.gz" contains the strong and 4F sector
\begin{eqnarray}
{\cal L}_{S4F} = {\cal L}_{S} +{\cal L}_{4fu}+
{\cal L}_{4fc}
\label{eqn:package3}
\end{eqnarray}
where the 4F lagrangians were presented in equations~\ref{eq:4Fu} and \ref{eq:4Fc}.
The relation between the parameters in equations~\ref{eq:4Fu} and \ref{eq:4Fc} and
the corresponding $co_i$ parameters in the param.dat file is shown in
table~\ref{tbl:dicLS4F}:
\begin{table}[h!]
\begin{center}
    \begin{tabular}{  c | c | c  }
    \hline \hline
	 $co27 \rightarrow  \alpha_{qq}^{1113}+\alpha_{qq'}^{1113}$  &  $co33 \rightarrow  \alpha_{qq}^{1123}+\alpha_{qq'}^{2113}$  &   $co39 \rightarrow \alpha_{uu}^{1123}$  \\
    \hline
	  $co28 \rightarrow  \alpha_{qu'}^{1311}$ & $co34 \rightarrow \alpha_{qq}^{2113}+\alpha_{qq'}^{1123}$ & $co40\rightarrow \alpha_{uu}^{2113}$  \\
    \hline
	  $co29 \rightarrow \alpha_{qu'}^{1113}$  & $co35 \rightarrow \alpha_{qu'}^{1321}$ & $co41 \rightarrow \alpha_{qu'}^{2311}$  \\
    \hline
	  $co30 \rightarrow \alpha_{uu}^{1113}$ & $co36 \rightarrow \alpha_{qu}^{1321}$ & $co42 \rightarrow \alpha_{qu'}^{1123}$ \\
    \hline
	 $co31 \rightarrow \alpha_{qu}^{1311}$  & $co37 \rightarrow \alpha_{qu'}^{2113}$ & $co43 \rightarrow \alpha_{qu}^{2311}$\\
    \hline
	 $co32 \rightarrow \alpha_{qu}^{1113}$  &  $co38 \rightarrow \alpha_{qu}^{2113}$  &  $co44 \rightarrow \alpha_{qu}^{1123}$ \\
    \hline \hline
    \end{tabular}
\end{center}
\caption{Coefficient dictionary for ${\cal L}_{S4F}$  }
\label{tbl:dicLS4F}
\end{table}

Finally we note that any combination of parameters can be made in a new package
and can be made available upon request. Generator and the different packages can be downloaded at
\textbf{http://coimbra.lip.pt/$\sim$miguelwon/MEtop/}.

\end{document}